\documentclass[10pt,journal,final]{IEEEtran}
\IEEEoverridecommandlockouts

\usepackage{amsfonts,amssymb}
\usepackage{ifpdf}
\usepackage{cite}
\usepackage{stfloats}
\usepackage{bm}
\usepackage{color,xcolor}
\usepackage{subfigure}
\ifCLASSINFOpdf
\usepackage[pdftex]{graphicx}
\graphicspath{{../pdf/}{../jpeg/}}
\DeclareGraphicsExtensions{.pdf,.jpeg,.png}
\else
\usepackage[dvips]{graphicx}

\graphicspath{{../eps/}}

\DeclareGraphicsExtensions{.eps}
\fi

\usepackage[cmex10]{amsmath}

\usepackage{algorithm}
\usepackage{algorithmic}
\ifCLASSOPTIONcompsoc
\else
\usepackage[caption=false,font=footnotesize]{subfig}
\fi

\usepackage{makecell}

\begin{document}
%\title{Joint Transmit Beamforming and IRS Reflection  Design  for  IRS-aided Secure  Integrated Sensing and Communication}	
\title{Secure  Intelligent Reflecting Surface Aided   Integrated Sensing and Communication}
	
\author{Meng~Hua,
	Qingqing~Wu,~\IEEEmembership{Senior Member,~IEEE,}
	  Wen~Chen,~\IEEEmembership{Senior Member,~IEEE,}
%Luxi~Yang,~\IEEEmembership{Senior Member,~IEEE,}
%Qingqing~Wu,
%Cunhua~Pan,
%Chunguo~Li,~\IEEEmembership{Senior Member,~IEEE,}
~Octavia A. Dobre,~\IEEEmembership{Fellow,~IEEE,}
and~A. Lee Swindlehurst,~\IEEEmembership{Fellow,~IEEE}

\thanks{M. Hua and Q. Wu are with the State Key Laboratory of Internet of Things for Smart City, University of Macau, Macao 999078, China (email: menghua@um.edu.mo; qingqingwu@um.edu.mo). }
\thanks{  W. Chen is with the Department of Electronic Engineering, Shanghai Institute of Advanced Communications and Data Sciences,
	Shanghai Jiao Tong University, Minhang 200240, China (e-mail:wenchen@sjtu.edu.cn).}
\thanks{ O. A. Dobre is with the Faculty of Engineering and Applied
	Science, Memorial University, St. John’s, NL A1B 3X5, Canada  (email: odobre@mun.ca).}
\thanks{A. L. Swindlehurst is with the Center for Pervasive Communications and Computing, University of California at Irvine, Irvine, CA 92697 USA (e-mail: swindle@uci.edu).}
}
\maketitle
%\vspace{-10cm}
\begin{abstract}	
In this paper, an  intelligent reflecting surface (IRS)
is  leveraged  to enhance the physical layer security of an integrated sensing and communication (ISAC) system  in which the IRS is deployed to not only assist the downlink communication for multiple users, but also create a  virtual line-of-sight (LoS) link for  target sensing.  In particular, we consider a challenging scenario where the target may be a  suspicious eavesdropper that potentially intercepts the communication-user information transmitted by the  base station (BS). To ensure
the sensing quality while preventing the eavesdropping,  dedicated sensing signals are transmitted by the BS. We investigate the joint design of the phase shifts at the IRS and the communication as well as radar beamformers at the BS to 
maximize the sensing beampattern gain towards the target, subject to the  maximum
information leakage to the  eavesdropping target and   the minimum signal-to-interference-plus-noise ratio (SINR) required  by users. 
Based on  the availability of  perfect channel state information (CSI) of all involved user links and the  accurate  target location  at the BS, two scenarios are considered and two different optimization algorithms are proposed. For the ideal scenario where the CSI of the user links and the   target location are perfectly known at the 	BS, a penalty-based algorithm is proposed to obtain a high-quality solution. In particular, the beamformers are obtained  with  a semi-closed-form solution   using  Lagrange duality   and the IRS phase shifts are solved for in closed form  by applying the majorization-minimization (MM) method.  On the other hand, for the more    practical scenario where the CSI is imperfect and the target location is uncertain,  a robust     algorithm  based on  the $\cal S$-procedure and sign-definiteness approaches is proposed.  Simulation results demonstrate the effectiveness of the proposed scheme in achieving a  trade-off between the communication quality  and the  sensing quality, and  also show the tremendous potential of IRS for use in sensing  and  improving the security  of ISAC systems.
% 
%the effectiveness of the proposed algorithms and  
%
%significantly improve the target sensing quality  while guaranteeing a desired level of secrecy for the  ISAC system.
\end{abstract}

\section{Introduction}
Driven by  emerging applications  for high-accuracy sensing services such as autonomous driving,  robot navigation, and intelligent traffic monitoring, etc.,  a new paradigm  is required to shift from   communication-based network designs  to networks with sensing-communication integration \cite{Liu2022survey}.  The research on the integration of  sensing and communication  networks  has recently attracted significant attention along the following two directions: radar-communication coexistence  \cite{zheng2019Radar} and integrated sensing and communication (ISAC)  \cite{liu2022integrated}. In the former,  the radar transceiver and the
communication transmitter are geographically separated, which usually  results in strong co-channel interference and requires prohibitive feedback  overhead  to  exchange information for coordination between two systems. For the latter,  the radar and communication functionalities share a common  hardware platform, which  leads to   both  integration   and coordination gains.

Recently, we are witnessing a booming interest from both academia and industry on  ISAC   systems due to their reduced  hardware cost, lower power consumption, and more efficient radio spectrum usage \cite{zhang2022enabling}.  Based on  design priorities and   underlying requirements, ISAC systems   can be   classified into three categories: communication-centric (C\&C) designs \cite{Sturm2011waveform}, radar-centric (R\&C)  designs \cite{Hassanien2016dual}, and joint  waveform designs \cite{joint2018liu,liu2018towards,hua2021optimal}. For   C\&C design,  the sensing functionality is integrated into the existing communication platform, where the communication performance has the highest priority.  The  objective  of this type of design is to  exploit the communication waveform  to implement the sensing functionality while  satisfying the quality-of-service (QoS) of the communication users. In contrast to  the   C\&C design,   
sensing  has the highest priority in  R\&C designs. The  objective  of this  approach is to  modulate the information into the sensing waveform to realize the communication functionality    without significantly degrading the sensing performance. The performance of the two types of designs above is  fundamentally limited by the hardware platforms and signal processing algorithms and fails to achieve a scalable tradeoff between sensing and communication.  The last  category, i.e., joint waveform design,    creates new waveforms instead of relying  on   existing communication or radar waveforms,  and provides additional degrees of freedom (DoFs) to support high data rates and to  improve sensing quality. As an example of the joint design approach,  the authors in \cite{joint2018liu}   revealed  that   communication-only waveform design is inferior to the joint design of communication   and radar waveforms in terms of beampattern synthesis, especially when the number of communication users is less than the number of targets. However, the ISAC system performance   is  significantly deteriorated by   unfavorable propagation environments with signal blockages, especially for   target sensing. In general, only   the reflected echo signals   that pass through  line-of-sight (LoS) links   are treated as useful information for  sensing while  non-LoS (NLoS) links are  treated as harmful interference or clutter. 
Unmanned aerial vehicles (UAVs) have been leveraged to assist    ISAC systems    since the UAV can establish strong LoS links between the UAV and users/targets by adjusting its trajectory  or deployment \cite{meng2022throughput,lyu2021joint,Meng2022UAV,9456851}. However, the UAV-enabled ISAC is not suitable for providing long-term coverage  due to  the  inherently limited battery capacity available on a UAV.   This raises a new open  question: How to  provide long-term and ubiquitous sensing coverage in harsh environments where the channel links are blocked in the ISAC system?  

Recently,  intelligent reflecting surface (IRS) technology has attracted significant attention and is  regarded as a promising    technology towards for beyond-fifth-generation (B$5\rm $G) and  sixth-generation ($6\rm $G) systems,   due to its capability of manipulating  the  wireless  propagation environment   with low power consumption and hardware cost \cite{WU2020towards,pan2021overview}.  Specifically, an IRS is a two-dimensional planar array
comprising a large number of  sub-wavelength metallic units, each of which is able to independently control  the phases and/or amplitudes of  impinging signals. Due to the small size of
each reflecting unit, a reasonably-sized IRS can be constructed with a large number of reflecting elements  and can   provide significant beamforming gains to  compensate for signal attenuation over  
long distances \cite{wu2020Beamforming}.   Motivated by this, IRS technology has   been extensively
investigated in the literature for various applications such as  mobile edge computing (MEC) \cite{chen2021irsaided,Bai2020latency,chen2022active}, wireless power transfer \cite{wu2020jointActive,chu2021novel,wu2021irsxxx,hua2021jointdynamic}, and multi-cell cooperation \cite{pan2020multicell,xie2021maxmin,hua2020intelligent}. The use of IRS is very appealing for ISAC since it is able to create virtual LoS links for both communication and sensing. Some representative works, see e.g.,  \cite{Buzzi2022Foundations,Lu2021Target,shao2022target}, have  studied the use  of IRS for sensing    and  verified their potential for enhancing   target sensing.  A handful of related works have been conducted on IRS-aided ISAC in the literature, see \cite{9364358,song2021joint,9416177,liu2021joint,sankar2022beamforming}, via jointly optimizing IRS phase  shifts and transmit beamformers to increase the sensing quality while satisfying communication QoS of the users.  However, the  above works assumed that the targets are  not attempting to intercept the transmitted signals. In   ISAC systems, the transmitted  signals may not only contain sensing signals but also communication signals,   which could be   readily intercepted by malicious targets. The problem of maintaining the communication QoS and the target sensing performance while also ensuring limited information leakage to   the targets has received very little attention.
Although works \cite{su2020secure} and  \cite{9737364} studied   secure transmission designs for ISAC system, the role of  IRS for sensing and communication was not unveiled 
 and  their proposed transceiver designs are also no longer applicable in the presence  of an IRS. 

%Although \cite{su2020secure} and  \cite{9737364} studied secure transmission designs for ISAC systems, they did not exploit the availability of IRS to assist in the process and  the  transceiver design was no longer applicable to the ISAC system with the availability of IRS. 
%Note also that the prior work has focused on beamforming design for cases where perfect channel state information (CSI) is available for all users at the BS, and ISAC designs for scenarios with imperfect CSI is still an open problem.

%Note that  all  above works  studied the beamforming  design by assuming that    perfect channel state information (CSI)  of all involved communication user links is available  at the base station (BS).  For the imperfect  CSI  scenario, how to design robust  secure transmission in IRS-aided ISAC system is still unknown. We note that
%although works \cite{su2020secure} and  \cite{9737364} studied the secure transmission design in ISAC system, the role of  IRS for sensing and communication is not unveiled 
% and  the  transceiver design is no longer applicable to the ISAC system with IRS. 

Motivated by the above  issues,  in this paper  we study a  secure IRS-aided  ISAC system  where the IRS is leveraged to not only assist the downlink communication from the base station (BS) to  multiple legitimate  users, but to also create a  virtual LoS link for  target sensing.  In addition, we consider a challenging scenario where the target may be  an eavesdropper that desires to intercept    information transmitted by the  BS. The  main contributions of this paper
are summarized as follows:
\begin{itemize}
	\item We study an IRS-aided ISAC system for enhancing the physical layer security  and realizing both communication and sensing.  To ensure
	the sensing quality while preventing   eavesdropping,   dedicated sensing signals are transmitted at the BS. Our objective  is to maximize the sensing beampattern gain by jointly optimizing the  communication beamformers, radar beamformers, and IRS phase shifts, subject to the maximum tolerable  information leakage to the  eavesdropping target and  the minimum signal-to-interference-plus-noise ratio (SINR) required  by the users. 
%	Two    different algorithms, i.e., the  penalty-based   algorithm and  the  robust design algorithm, are  proposed to cater to two different optimization problems.	
	Based  on whether or not   perfect channel state information (CSI) and     target location information are known by BS,   two different optimization problems are formulated. Subsequently, two    different algorithms are  proposed, i.e., a  penalty-based   algorithm and  a robust   algorithm.	
	\item  For the ideal scenario where the CSI of the  user links and the   target location are known at the 	BS, the resulting optimization problem is non-convex due to the presence of coupled optimization variables in both the  objective function and  constraints. In addition, the unit-modulus constraint imposed on each IRS phase shift   renders the  formulated problem more difficult to solve.  To address this difficulty, a penalty-based algorithm is proposed in which the beamformers are obtained  with  a semi-closed-form solution  using  Lagrange duality  and the IRS phase shifts are obtained with a closed-form solution by applying   majorization-minimization (MM), both of which significantly reduce the computational  complexity of the penalty-based  algorithm.
	
	\item For the more practical scenario where   perfect CSI of   communication channels and the  target location are not available  at the BS, we design a robust transmission strategy. The resulting optimization problem involving an infinite number of constraints  is more challenging  to solve than the former one, and the previous penalty-based algorithm   is  no longer  applicable. To solve this optimization problem, the $\cal S$-procedure and sign-definiteness approaches are applied to transform the   infinite number of inequalities into a finite  number of 
	linear matrix inequalities (LMIs). Then, an efficient alternating optimization (AO) algorithm is   proposed that toggles between optimizing the transmit beamformers and IRS phase shifts.

	\item Our simulation results verify  the effectiveness of the proposed scheme in achieving a  trade-off between the communication quality  and the target  sensing quality and validate the tremendous potential of IRS to achieve significant beampattern gains and  guarantee  ISAC system security. Our results also  show that   dedicated sensing signals are required to  further improve the system performance. 
\end{itemize} 
The rest of the paper is organized as follows. Section II introduces the system model and problem formulations  for the considered IRS-aided secure ISAC system. In Section III, a penalty-based algorithm is proposed to solve the perfect CSI and the known-target location case.  In Section IV, a robust design algorithm  is proposed to solve the case with imperfect CSI and  uncertain  target location. Numerical results are provided in Section V and the paper is concluded in Section VI.

\emph{Notations}: Boldface upper-case and lower-case  letters denote matrices and vectors,  respectively.  ${\mathbb C}^ {d_1\times d_2}$ stands for the set of  complex $d_1\times d_2$  matrices. For a complex-valued vector $\bf x$, ${\left\| {\bf x} \right\|}$ represents the  Euclidean norm of $\bf x$, ${\rm arg}({\bf x})$ denotes  a vector containing the phase of the elements of   $\bf x$, and ${\rm diag}(\bf x) $ denotes a diagonal matrix whose   diagonal elements are given by the elements of  $\bf x$.
${\left(  \cdot  \right)^T},{\left(  \cdot  \right)^*}$, and ${\left(  \cdot  \right)^H}$  stand for  the transpose operator,  conjugate operator, and conjugate transpose operator, respectively.  ${\left\| {\bf{X}} \right\|_F}$ and ${\rm{rank}}\left( {{{{\bf{X}}}}} \right)$  represent the  Frobenius norm and rank of ${\bf{X}}$, respectively,  and  ${\bf{X}} \succeq {\bf{0}}$ indicates that matrix $\bf X$ is positive semi-definite.
A circularly symmetric complex Gaussian random variable $x$ with mean $ \mu$ and variance  $ \sigma^2$ is denoted by ${x} \sim {\cal CN}\left( {{{\mu }},{{\sigma^2 }}} \right)$. $ \otimes $ denotes the Kronecker product operator and  ${\cal O}\left(  \cdot  \right)$ indicates the big-O computational complexity.
\begin{figure}[!t]
	\centerline{\includegraphics[width=3in]{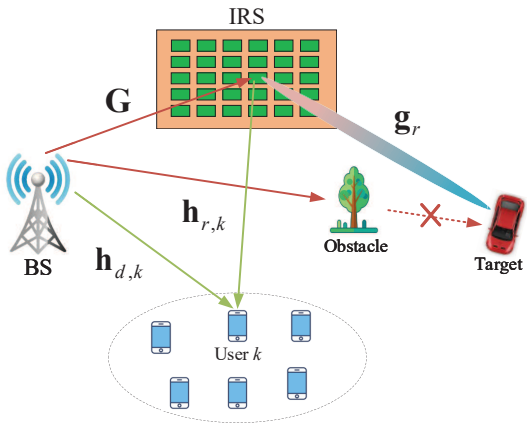}}
	\caption{An IRS-aided secure ISAC system.} \label{fig1}
%	\vspace{-0.6cm}
\end{figure}
\section{System Model and Problem Formulation}
\subsection{System Model}
We consider  a secure IRS-aided  ISAC system that comprises a dual-function  BS, an IRS, one target of interest,\footnote{Although we consider a single target in this work,   the   algorithms  proposed for the single-target case   are   applicable to the  multi-target case with only slight modifications.} and $K$ single-antenna  users, as shown in Fig.~\ref{fig1}.  The BS is equipped with a uniform linear  array  with $N$ transmit antennas, and the IRS is a uniform planar array with  $M$ reflecting elements. For
convenience, we denote the sets of users,  BS transmit antennas, and IRS reflecting elements as $\cal K$,  $\cal N$, and  $\cal M$, respectively.

We assume that  both information signals and radar signals are simultaneously transmitted  for communication and sensing.  The transmitted signals at the  BS can be expressed as 
  \begin{align}
{\bf{s}} = \sum\limits_{k = 1}^K {{{\bf{w}}_{c,k}}{x_{c,k}}}  + \sum\limits_{n = 1}^N {{{\bf{w}}_{r,n}}{x_{r,n}}} , \label{transmittedssignal}
\end{align}
where $x_{c,k}$ denotes the information signal for user $k$ assumed to satisfy ${x_{c,k}} \sim {\cal CN}\left( {0,1} \right)$ and ${{\bf{w}}_{c,k}} \in {{\mathbb C}^{N \times 1}}$ represents its corresponding communication  beamformer. Similarly,  $x_{r,n}$ denotes the $n$th radar signal satisfying ${\mathbb E}\left\{ {{x_{r,n}}} \right\} = 0$ and  ${\mathbb E}\left\{ {{{\left| {{x_{r,n}}} \right|}^2}} \right\} = 1$,   and ${{\bf{w}}_{r,n}} \in {{\mathbb C}^{N \times 1}}$ represents the corresponding radar  beamformer. We assume that   communication
and radar signals are statistically independent and uncorrelated, i.e., ${\mathbb E}\left\{ {{x_{r,n}}x_{c,k}^H} \right\} = 0, \forall k,n$.

\subsubsection{Communication Model} We consider quasi-static block-fading channels and focus on a given fading
  block during which all the channels involved are assumed to remain unchanged. Let 
  ${\bf{G}}\in {{\mathbb C}^{M \times N}}$ denote the complex equivalent baseband   channel from the BS to the IRS, 
 ${\bf{h}}_{r,k}^H \in {{\mathbb C}^{1 \times M}}$ denote that from the IRS to user $k$, 
 and ${{\bf h}^H_{d,k}} \in {{\mathbb C}^{1 \times N}}$ denote that from the BS to user $k$, $k\in{\cal K}$.  
% We assume that the CSI of all involved channels, i.e., ${\bf{G}}$, ${\rm{diag}}\left( {{\bf{h}}_{r,k}^H} \right){\bf{G}}$, and ${\bf{h}}_{d,k}^H$,  is available at the  BS  by applying the state-of-the-art two-timescale channel estimation method \cite{9400843}. 
% Note that  the CSI of  the BS-IRS link, i.e., ${\bf{G}}$, is required in this paper for target sensing, which will be shown  later. 
 We assume that the CSI of all involved channels, i.e., ${\bf{G}}$, ${\rm{diag}}\left( {{\bf{h}}_{r,k}^H} \right){\bf{G}}$, and ${\bf{h}}_{d,k}^H$,  is available at the  BS  by applying  channel estimation methods, e.g., \cite{9400843}. 
%Note that  the CSI of  the BS-IRS link, i.e., ${\bf{G}}$, is required in this paper for target sensing, which will be shown  later. 
 The signal received at user $k$ is given by 
 \begin{align}
{y_k} = \left( {{\bf{h}}_{r,k}^H{\bf{\Theta G}} + {{\bf{h}}_{d,k}^H}} \right){\bf{s}} + {n_k}, k \in {\cal K},
 \end{align}
 where ${\bf{\Theta }} = {\rm{diag}}\left( {{v_1}, \ldots ,{v_M}} \right)$ represents the IRS reflection phase shift matrix and ${n_k} \sim {\cal CN}\left( {0,\sigma _k^2} \right)$ denotes the noise received at user $k$. Accordingly, the received SINR at user $k$  is given by
  \begin{align}
{\gamma _k} = \frac{{{{\left| {{\bf{h}}_k^H{{\bf{w}}_{c,k}}} \right|}^2}}}{{\sum\limits_{i \ne k}^K {{{\left| {{\bf{h}}_k^H{{\bf{w}}_{c,i}}} \right|}^2}}  + \sum\limits_{n = 1}^N {{{\left| {{\bf{h}}_k^H{{\bf{w}}_{r,n}}} \right|}^2}}  + \sigma _k^2}},  k \in {\cal K},
\end{align}
where ${\bf{h}}_k^H = {\bf{h}}_{r,k}^H{\bf{\Theta G}} + {{\bf{h}}_{d,k}^H}$. 
\subsubsection{Radar Sensing and Interception Model} We consider the   scenario where the direct link between the BS and the potential target is not available due to the blockages. To tackle this issue, the IRS is leveraged to create a  virtual LoS link between the IRS and the target,  thereby establishing an effective BS-IRS-target link  for sensing.  Let $\theta $ and $\varphi $ denote the azimuth and elevation angle-of-departure (AoD) from  the IRS to the target, respectively. Accordingly,  the steering vector from  the IRS to  the target at  direction $\left( {\theta ,\varphi } \right)$ can be expressed as 
\begin{align}
{\bf{g}}_r^H& =  \alpha_r\left[ {1,{e^{ - j\frac{{2\pi d}}{\lambda }\sin \theta \cos \varphi }}, \ldots {e^{ - j\frac{{2\pi \left( {{M_x} - 1} \right)d}}{\lambda }\sin \theta \cos \varphi }}} \right]  \notag\\
&\otimes\left[ {1,{e^{ - j\frac{{2\pi d}}{\lambda }\sin \theta \sin \varphi }}, \ldots {e^{ - j\frac{{2\pi \left( {{M_z} - 1} \right)d}}{\lambda }\sin \theta \sin \varphi }}} \right], \label{IRS_target_steervector}
\end{align} 
where $M_x$ and $M_z$ denote the numbers  of reflecting elements along $x$-axis and $z$-axis, respectively,  ${\alpha _r} $ represents the large-scale fading  coefficient, $\lambda $ denotes the wavelength, and $d$ denotes the spacing
between two adjacent reflecting elements.  The received signal at the target  is given by 
\begin{align}
{y_t} = {\bf{g}}_r^H{\bf{\Theta G}}\left( {\sum\limits_{k = 1}^K {{{\bf{w}}_{c,k}}{x_{c,k}}}  + \sum\limits_{n = 1}^N {{{\bf{w}}_{r,n}}{x_{r,n}}} } \right) + {n_t},
\end{align}
where ${n_t} \sim {\cal CN}\left( {0,\sigma _t^2} \right)$ represents the  noise received at  the target.
As a result, the beampattern
gain  towards the target is given by 
\begin{align}
{\cal P} & = {\mathbb E}\left\{ {{{\left| { {{\bf{g}}^H}\left( {\sum\limits_{k = 1}^K {{{\bf{w}}_{c,k}}{x_{c,k}}}  + \sum\limits_{n = 1}^N {{{\bf{w}}_{r,n}}{x_{r,n}}}} \right)} \right|}^2}} \right\}\notag \\
&={{\bf{g}}^H}\left( {\sum\limits_{k = 1}^K  {{{\bf{w}}_{c,k}}{\bf{w}}_{c,k}^H}  + \sum\limits_{n = 1}^N {{{\bf{w}}_{r,n}}{\bf{w}}_{r,n}^H} } \right){\bf{g}},
\end{align}
 where ${{\bf{g}}^H} = {\bf{g}}_r^H{\bf{\Theta G}}$. 
 
Since the target is  a potential eavesdropper, it tries to decode
 information from its received signals.    
 The  SINR received at the target for intercepting  user $k$'s information is  given by 
   \begin{align}
{\gamma _{e,k}} = \frac{{{{\left| {{{\bf{g}}^H}{{\bf{w}}_{c,k}}} \right|}^2}}}{{\sum\limits_{i \ne k}^K {{{\left| {{{\bf{g}}^H}{{\bf{w}}_{c,i}}} \right|}^2}}  + \sum\limits_{n = 1}^N {{{\left| {{{\bf{g}}^H}{{\bf{w}}_{r,n}}} \right|}^2}}  + \sigma _t^2}}, k \in {\cal K}. \label{SINR_leakage}
 \end{align}
\subsection{Problem Formulation}
The objective of this paper is to maximize the beampattern gain at the target by jointly optimizing the transmit  beamformers and IRS phase shifts, subject to the minimum SINR required by users and the maximum tolerable information leakage to the 	target. Depending on whether  perfect CSI of the communication channels and the   target location are available at the BS, we consider two scenarios elaborated as below.
\subsubsection{Perfect CSI and Known Target Location Scenario}In this scenario,    perfect CSI of the  communication channels  and the   target location are   known at the BS. Accordingly, the problem is formulated as 
 \begin{subequations} \label{type_I_P}
	\begin{align}
 &\mathop {\max }\limits_{\left\{ {{{\bf{w}}_{c,k}},{{\bf{w}}_{r,n}},{v_m}} \right\}} {{\bf{g}}^H}\left( {\sum\limits_{k = 1}^K {{{\bf{w}}_{c,k}}{\bf{w}}_{c,k}^H}  + \sum\limits_{n = 1}^N {{{\bf{w}}_{r,n}}{\bf{w}}_{r,n}^H} } \right){\bf{g}} \label{type_I_P_obj}\\
	& {\rm s.t.}~\gamma _k \ge {r_{k,{\rm{th}}}},k\in {\cal K},\label{type_I_P_const1}\\
	& \qquad{\gamma _{e,k}} \le {r_{e,k,{\rm{th}}}},k\in {\cal K},\label{type_I_P_const2}\\
	&\qquad \sum\limits_{k = 1}^K {{{\left\| {{{\bf{w}}_{c,k}}} \right\|}^2}}  + \sum\limits_{n = 1}^N {{{\left\| {{{\bf{w}}_{r,n}}} \right\|}^2}}  \le {P_{\max }},\label{type_I_P_const3}\\
	&\qquad \left| {{v_m}} \right| = 1, m\in {\cal M},\label{type_I_P_const4}
	\end{align}
\end{subequations}
where ${r_{k,{\rm{th}}}}$ in  \eqref{type_I_P_const1}  denotes the minimum SINR  required by user $k$,  ${r_{e,k,{\rm{th}}}}$ in  \eqref{type_I_P_const2} represents the  maximum
tolerable leakage of user $k$'s information
 to the target, $P_{\max}$ in  \eqref{type_I_P_const3}  stands for the maximum allowed transmit power at the BS, and constraint \eqref{type_I_P_const4} denotes the unit-modulus constraint imposed on each IRS phase shift. Note that with constraints  \eqref{type_I_P_const1}  and \eqref{type_I_P_const2}, the   level of  physical layer security of the ISAC system is guaranteed \cite{9133130}.

\subsubsection{Imperfect CSI and Uncertain  Target Location Scenario} In this scenario,   perfect CSI of  the communication channels is  not available at the BS, and   the precise location of the target is unknown but the region of  interest for sensing is available, i.e., ${\Phi _h} = \left[ {\theta  - \Delta \theta,  \theta  + \Delta \theta } \right]$ and ${\Phi _v} = \left[ {\varphi  - \Delta \varphi,  \varphi  + \Delta \varphi } \right]$ are known, where ${\Delta \theta }$ and ${\Delta \varphi {\kern 1pt} }$ represent the  azimuth and vertical  sensing range, respectively. Defining ${{\bf{F}}_k}{\rm{ = diag}}\left( {{\bf{h}}_{r,k}^H} \right){\bf{G}}$ and ${{\bf{F}}_r}{\rm{ = diag}}\left( {{\bf{g}}_r^H} \right){\bf{G}}$, we can rewrite  ${\bf{h}}_k^H = {\bf{h}}_{r,k}^H{\bf{\Theta G}} + {\bf{h}}_{d,k}^H{\rm{ = }}{{\bf{v}}^H}{{\bf{F}}_k} + {\bf{h}}_{d,k}^H$ and ${{\bf{g}}^H} = {\bf{g}}_r^H{\bf{\Theta G}} = {{\bf{v}}^H}{{\bf{F}}_r}$, where  ${{\bf{v}}^H} = \left[ {{v_1}, \ldots ,{v_M}} \right]$.
 The bounded CSI error models for channels ${{\bf{F}}_k}$, ${{\bf{F}}_r}$, and ${{\bf{h}}_{d,k}}$ are  respectively given by \cite{zhou2020framework}\footnote{The bounded CSI error models  for  ${{\bf{F}}_r}$ and  ${\bf{G}}$ are equivalent since ${{\bf{g}}_r^H}$ is a deterministic  LoS channel. For notational simplicity, we use  the bounded CSI error model  for  ${{\bf{F}}_r}$ in the sequel.}
\begin{align}
&{{\bf{F}}_k}{\rm{ = }}{{{\bf{\hat F}}}_k} \!+\! \Delta {{\bf {F}}_k}, \text{ with}~ {{\bf{\cal F}}_k}\! =\! \left\{ {\Delta {{\bf{F}}_k}\!:\!{{\left\| {\Delta {{\bf{F}}_k}} \right\|}_F} \!\le\! {\varepsilon _k}} \right\}, \forall k,\\
&{{\bf{F}}_r}{\rm{ = }}{{{\bf{\hat F}}}_r} + \Delta {{\bf{F}}_r},  \text{ with}~{{\bf{\cal F}}_r} = \left\{ {\Delta {{\bf{F}}_r}:{{\left\| {\Delta {{\bf{F}}_r}} \right\|}_F} \le {\varepsilon _r}} \right\},\\
&{\bf{h}}_{d,k} = {\bf{\hat h}}_{d,k} + \Delta {\bf{h}}_{d,k},\text{ with}~{{\bf{\cal H}}_{d,k}} = \left\{ {\Delta {{\bf{h}}_{d,k}}:\left\| {\Delta {{\bf{h}}_{d,k}}} \right\|} \right. \notag\\
&\qquad\qquad\qquad\qquad\qquad\qquad\qquad\quad \left. { \le {\varepsilon _{d,k}}} \right\},\forall k,
\end{align}
where ${\bf{\hat G}}$ represents the  estimated channel for the BS-IRS link,  ${{\bf{\hat F}}_k}$ denotes the estimated cascaded channel for  user $k$, ${{{\bf{\hat F}}}_r}{\rm{ = diag}}\left( {{\bf{g}}_r^H} \right){\bf{\hat G}}$ stands for the estimated cascaded channel for the target. 
Accordingly, the problem is formulated as\footnote{In this scenario,  we drop the direction indices, i.e., ${\theta _h}$ and ${\varphi _v}$, and use the notation ${{\bf{g}}}$ to represent ${{\bf{g}}}\left( {{\theta _h},{\varphi _v}} \right)$ for the brevity.}
  \begin{subequations} \label{Imperfect_P}
	\begin{align}
	&\mathop {\max }\limits_{\left\{ {{{\bf{w}}_{c,k}},{{\bf{w}}_{r,n}},{v_m}} \right\}} \mathop {\min }\limits_{\scriptstyle{\theta _h} \in {\Phi _h},\hfill\atop
		\scriptstyle{\varphi _v} \in {\Phi _v}\hfill} {{\bf{g}}^H}\left( {\sum\limits_{k = 1}^K {{{\bf{w}}_{c,k}}{\bf{w}}_{c,k}^H}  + \sum\limits_{n = 1}^N {{{\bf{w}}_{r,n}}{\bf{w}}_{r,n}^H} } \right){\bf{g}}\label{Imperfect_P_obj}\\
	& {\rm s.t.}~\gamma _k \ge {r_{k,{\rm{th}}}},\Delta {{\bf{h}}_{d,k}} \in  {{\bf{\cal H}}_{d,k}}, \Delta {{\bf{F}}_k} \in {{\bf{\cal F}}_k}, k\in {\cal K},\label{Imperfect_const1}\\
	& \quad{\gamma _{e,k}} \le {r_{e,k,{\rm{th}}}},{{\theta _h} \in {\Phi _h},{\varphi _v} \in {\Phi _v}}, \Delta {{\bf{F}}_r} \in  {{\bf{\cal F}}_r}, k\in {\cal K},\label{Imperfect_const2}\\
	&\quad \sum\limits_{k = 1}^K {{{\left\| {{{\bf{w}}_{c,k}}} \right\|}^2}}  + \sum\limits_{n = 1}^N {{{\left\| {{{\bf{w}}_{r,n}}} \right\|}^2}}  \le {P_{\max }},\label{Imperfect_const3}\\
	&\quad \left| {{v_m}} \right| = 1, m\in {\cal M}.\label{Imperfect_const4}
	\end{align}
\end{subequations}
The above two problems \eqref{type_I_P}  and \eqref{Imperfect_P} are both non-convex
due to the fact that the IRS reflection coefficients are constrained to be unit modulus, and because the optimization variables are  coupled
in both the  objective functions and constraints. In general, there     are no standard methods for solving such non-convex optimization problems optimally.   In particular, \eqref{Imperfect_const1} and \eqref{Imperfect_const2} involve an infinite number of inequalities, which makes problem \eqref{Imperfect_P} even more difficult to address.   In the following, we first propose a penalty-based algorithm for solving problem \eqref{type_I_P} in Section III and then propose an AO algorithm based on the $\cal S$-procedure and sign-definiteness approaches for solving problem \eqref{Imperfect_P} in Section IV.  
\section{Proposed Solution for Perfect CSI and Known Target Location}
 In this section, we consider the case where     perfect CSI and    target location are known at the BS, which   provides a performance upper bound for the case with imperfect CSI and uncertain target location. To obtain a high-quality solution for problem \eqref{type_I_P}, a penalty-based algorithm is proposed to decouple constraint coupling between the optimization  variables in different blocks. Define auxiliary variables $\left\{ {{y_{c,k}},{y_{r,n}},{z_{c,k,i}},{z_{r,k,n}}},i \in {\cal K},k \in {\cal K},n \in {\cal N} \right\}$ and let  
${{\bf{g}}^H}{{\bf{w}}_{c,k}}{\rm{ = }}{y_{c,k}},{{\bf{g}}^H}{{\bf{w}}_{r,n}} = {y_{r,n}},{\bf{h}}_k^H{{\bf{w}}_{c,i}} = {z_{c,k,i}}$, and ${\bf{h}}_k^H{{\bf{w}}_{r,n}} = {z_{r,k,n}}$.
Problem \eqref{type_I_P} can be equivalently transformed  as 
\begin{subequations}\label{type_I_P_new}
\begin{align}
&\mathop {\max }\limits_{\left\{ {{{\bf{w}}_{c,k}}} \right\},\left\{ {{{\bf{w}}_{r,n}}} \right\},\left\{ {{v_m}} \right\},\Omega } \sum\nolimits_{k = 1}^K {{{\left| {{y_{c,k}}} \right|}^2}}  + \sum\nolimits_{n = 1}^N {{{\left| {{y_{r,n}}} \right|}^2}}\label{type_I_P_new_obj} \\
&{\rm s.t.}~\frac{{{{\left| {{z_{c,k,k}}} \right|}^2}}}{{\sum\limits_{i \ne k}^K {{{\left| {{z_{c,k,i}}} \right|}^2} + \sum\limits_{n = 1}^N {{{\left| {{z_{r,k,n}}} \right|}^2} + \sigma _k^2} } }} \ge {r_{k,{\rm th}}},k \in {\cal K},\label{type_I_P_new_const1}\\
&\quad\frac{{{{\left| {{y_{c,k}}} \right|}^2}}}{{\sum\limits_{i \ne k}^K {{{\left| {{y_{c,i}}} \right|}^2}}  + \sum\limits_{n = 1}^N {{{\left| {{y_{r,n}}} \right|}^2}}  + \sigma _t^2}} \le {r_{e,k,{\rm{th}}}},k \in {\cal K},\label{type_I_P_new_const2}\\
& \quad {{\bf{g}}^H}{{\bf{w}}_{c,k}}{\rm{ = }}{y_{c,k}},{{\bf{g}}^H}{{\bf{w}}_{r,n}} = {y_{r,n}},{\bf{h}}_k^H{{\bf{w}}_{c,k}} = {z_{c,k,i}},\notag\\
&\quad\qquad \qquad  {\bf{h}}_k^H{{\bf{w}}_{r,n}} = {z_{r,k,n}},i \in {\cal K},k \in {\cal K},n \in {\cal N},\label{type_I_P_new_const3}\\
&\quad \eqref{type_I_P_const3}, \eqref{type_I_P_const4},
\end{align}
\end{subequations}
where $\Omega  = \left\{ {{y_{c,k}},{y_{r,n}},{z_{c,k,i}},{z_{r,k,n}}} \right\}$.
We then reformulate \eqref{type_I_P_new_const3} as penalty terms that are added to the objective function \eqref{type_I_P_new_obj}   yielding the following   optimization problem 
\begin{subequations}\label{type_I_P_new_penbalty}
\begin{align}
&\mathop {\max }\limits_{\left\{ {{{\bf{w}}_{c,k}},{{\bf{w}}_{r,n}},{v_m}} \right\},\Omega } \sum\nolimits_{k = 1}^K {{{\left| {{y_{c,k}}} \right|}^2}}  + \sum\nolimits_{n = 1}^N {{{\left| {{y_{r,n}}} \right|}^2}}  - \frac{1}{{2\rho }}\times \notag\\
&\left( {\sum\nolimits_{k = 1}^K {{{\left| {{{\bf{g}}^H}{{\bf{w}}_{c,k}} - {y_{c,k}}} \right|}^2} + } } \right.\sum\nolimits_{n = 1}^N {{{\left| {{{\bf{g}}^H}{{\bf{w}}_{r,n}} - {y_{r,n}}} \right|}^2} + } \notag\\
&\!\!\!\!\left. {\sum\limits_{k = 1}^K {\sum\limits_{i = 1}^K {{{\left| {{\bf{h}}_k^H{{\bf{w}}_{c,i}} \!-\! {z_{c,k,i}}} \right|}^2} \!+ \! \sum\limits_{k = 1}^K {\sum\limits_{n = 1}^N {{{\left| {{\bf{h}}_k^H{{\bf{w}}_{r,n}} \!-\! {z_{r,k,n}}} \right|}^2}} } } } } \right) \\
&{\rm s.t.}~\eqref{type_I_P_const3}, \eqref{type_I_P_const4}, \eqref{type_I_P_new_const1},\eqref{type_I_P_new_const2},
\end{align}
\end{subequations}
where $\rho>0$ represents the  parameter that penalizes the violations of the equality constraints in \eqref{type_I_P_new_const3}. To address problem \eqref{type_I_P_new_penbalty}, a penalty-based algorithm comprising two layers is proposed, where in the outer layer, we gradually update the  penalty parameter, while in the inner loop,  we  alternately optimize the   variables in different blocks.
\subsection{Inner Layer Optimization}
In the inner layer, we  divide all the optimization variables into three  blocks: 1) auxiliary variable set $\Omega $, 2) transmit beamformers $\left\{ {{{\bf{w}}_{c,k}},{{\bf{w}}_{r,n}}} \right\}$, and  3) IRS phase shifts  ${\left\{ {{v_m}} \right\}}$. 
%In particular, we obtain a closed-form and/or a semi-closed-form solution to each of these three subproblems.
\subsubsection{Optimizing $\Omega $  for given $\left\{ {{{\bf{w}}_{c,k}},{{\bf{w}}_{r,n}}} \right\}$  and ${\left\{ {{v_m}} \right\}}$}This subproblem can be written as
\begin{subequations}
	\begin{align}
&\mathop {\max }\limits_{\Omega } \sum\nolimits_{k = 1}^K {{{\left| {{y_{c,k}}} \right|}^2}}  + \sum\nolimits_{n = 1}^N {{{\left| {{y_{r,n}}} \right|}^2}}  - \frac{1}{{2\rho }}\times \notag\\
&\left( {\sum\nolimits_{k = 1}^K {{{\left| {{{\bf{g}}^H}{{\bf{w}}_{c,k}} - {y_{c,k}}} \right|}^2} + } } \right.\sum\nolimits_{n = 1}^N {{{\left| {{{\bf{g}}^H}{{\bf{w}}_{r,n}} - {y_{r,n}}} \right|}^2} + } \notag\\
&\!\!\!\!\! \left. {\sum\limits_{k = 1}^K {\sum\limits_{i = 1}^K {{{\left| {{\bf{h}}_k^H{{\bf{w}}_{c,i}} \!-\! {z_{c,k,i}}} \right|}^2} \!+ \! \sum\limits_{k = 1}^K {\sum\limits_{n = 1}^N {{{\left| {{\bf{h}}_k^H{{\bf{w}}_{r,n}} \!-\! {z_{r,k,n}}} \right|}^2}} } } } } \right) \\
		&{\rm s.t.}~\eqref{type_I_P_new_const1},\eqref{type_I_P_new_const2}.
	\end{align}
\end{subequations}
Since the optimization
variables with respect to (w.r.t.) different blocks $\left\{ {{y_{c,k}},{y_{r,n}},\forall k,\forall n} \right\}$ and $\left\{ {{z_{c,k,i}},{z_{r,k,n}},\forall i,\forall n} \right\}$ for $k \in {\cal K}$ are separable in both the objective function and
constraints, we can independently solve $K+1$ subproblems in parallel. Specifically, the subproblem corresponding to   the $k$th block  $\left\{ {{z_{c,k,i}},{z_{r,k,n}},\forall i,\forall n} \right\}$ is given by 
\begin{subequations} \label{auxiliary_sub1}
	\begin{align}
		&\!\!\!\mathop {\min }\limits_{\left\{ {\scriptstyle{z_{c,k,i}},\hfill\atop
				\scriptstyle{z_{r,k,n}}\hfill} \right\}} \sum\limits_{i = 1}^K {{{\left| {{\bf{h}}_k^H{{\bf{w}}_{c,i}}\! -\! {z_{c,k,i}}} \right|}^2} + } \sum\limits_{n = 1}^N {{{\left| {{\bf{h}}_k^H{{\bf{w}}_{r,n}}\! -\! {z_{r,k,n}}} \right|}^2}}   \\
		&{\rm s.t.}~\frac{{{{\left| {{z_{c,k,k}}} \right|}^2}}}{{\sum\limits_{i \ne k}^K {{{\left| {{z_{c,k,i}}} \right|}^2} + \sum\limits_{n = 1}^N {{{\left| {{z_{r,k,n}}} \right|}^2} + \sigma _k^2} } }} \ge {r_{k,{\rm th}}}.\label{type_I_P_new_const1_one}
	\end{align}
\end{subequations}
It is not difficult to see that problem \eqref{auxiliary_sub1} is a quadratically constrained
quadratic program (QCQP)
  with a convex objective function and non-convex constraint \eqref{type_I_P_new_const1_one}. Fortunately, it was shown
in \cite[Appendix B.1]{boyd2004convex}
that   strong duality holds for any optimization problem with a quadratic
objective and one quadratic inequality constraint, provided that the Slater’s condition holds. Therefore, we can solve problem \eqref{auxiliary_sub1} by solving its dual problem. Specifically, by introducing dual variable $\mu_1\ge0$ associated with constraint \eqref{type_I_P_new_const1_one}, the Lagrangian function
of problem \eqref{auxiliary_sub1} is given by
\begin{align}
	&{{\cal L}_1}\left( {{z_{c,k,i}},{z_{r,k,n}},{\mu _1}} \right) =\notag\\
	& \sum\limits_{i = 1}^K {{{\left| {{\bf{h}}_k^H{{\bf{w}}_{c,k}} - {z_{c,k,i}}} \right|}^2}}  + \sum\limits_{n = 1}^N {{{\left| {{\bf{h}}_k^H{{\bf{w}}_{r,n}} - {z_{r,k,n}}} \right|}^2}}+ {\mu _1}\times \notag\\
	&\!\!\!\!\!  \left( {{r_{k,{\rm{th}}}}\left( {\sum\limits_{i \ne k}^K {{{\left| {{z_{c,k,i}}} \right|}^2} + \sum\limits_{n = 1}^N {{{\left| {{z_{r,k,n}}} \right|}^2} + \sigma _k^2} } } \right) - {{\left| {{z_{c,k,k}}} \right|}^2}} \right). \label{larg_function_1}
\end{align}
Accordingly, the corresponding dual function is given by $\mathop {\min }\limits_{{z_{c,k,i}},{z_{r,k,n}}} {{\cal L}_1}\left( {{z_{c,k,i}},{z_{r,k,n}},{\mu _1}} \right)$. It can be readily checked  that to make  the dual function  bounded, we must have $0 \le {\mu _1} < 1$. Taking the first-order derivative of ${{\cal L}_1}\left( {{z_{c,k,i}},{z_{r,k,n}},{\mu _1}} \right)$  w.r.t. ${{z_{c,k,i}}}$ and ${{z_{r,k,n}}}$ and setting both  to zero, we   obtain the optimal solution as
\begin{align}
	&z_{c,k,i}^{{\rm{opt}}}\left( {{\mu _1}} \right) = \left\{ \begin{array}{l}
		\frac{{{\bf{h}}_k^H{{\bf{w}}_{c,i}}}}{{1 + {\mu _1}{r_{k,{\rm{th}}}}}},i \ne k, i\in {\cal K},\\
		\frac{{{\bf{h}}_k^H{{\bf{w}}_{c,k}}}}{{1 - {\mu _1}}},i = k,
	\end{array} \right.\label{larg_function_2} \\
	&z_{r,k,n}^{{\rm{opt}}}\left( {{\mu _1}} \right) = \frac{{{\bf{h}}_k^H{{\bf{w}}_{r,n}}}}{{1 + {\mu _1}{r_{k,{\rm{th}}}}}}, n \in {\cal N}. \label{larg_function_3}
\end{align}
If constraint \eqref{type_I_P_new_const1_one} is not met with equality at the optimal solution, i.e.,  $\mu _1^{{\rm{opt}}} = 0$, then the optimal solutions to problem \eqref{auxiliary_sub1} are  given by  $z_{c,k,i}^{{\rm{opt}}}\left( 0 \right)$ and $z_{r,k,n}^{{\rm{opt}}}\left( 0 \right)$. Otherwise, the optimal $\mu _1^{{\rm{opt}}}$ is a positive value ($0<\mu _1^{{\rm{opt}}}<1$) that satisfies the  equality constraint \eqref{type_I_P_new_const1_one}, i.e.,  
\begin{align}
	&{r_{k,{\rm{th}}}}\left( {\sum\limits_{i \ne k}^K {{{\left| {z_{c,k,i}^{{\rm{opt}}}\left( {\mu _1^{{\rm{opt}}}} \right)} \right|}^2} + \sum\limits_{n = 1}^N {{{\left| {z_{r,k,n}^{{\rm{opt}}}\left( {\mu _1^{{\rm{opt}}}} \right)} \right|}^2} + \sigma _k^2} } } \right) \notag\\
	&\qquad\qquad\qquad\qquad\qquad\qquad\quad -{\left| {z_{c,k,k}^{{\rm{opt}}}\left( {\mu _1^{{\rm{opt}}}} \right)} \right|^2} = 0.
\end{align}
It can be readily verified  that ${{{\left| {z_{c,k,i}^{{\rm{opt}}}\left( {{\mu _1}} \right)} \right|}^2}}$ for $i \ne k$ and $z_{r,k,n}^{{\rm{opt}}}\left( {{\mu _1}} \right)$ are
both  monotonically decreasing with ${{\mu _1}}$, while $z_{c,k,k}^{{\rm{opt}}}\left( {{\mu _1}} \right)$
is monotonically increasing with ${{\mu _1}}$ for $0<\mu _1<1$. As such, the optimal $\mu _1^{{\rm{opt}}}$  can be obtained by applying a simple bisection search method between $0$ and $1$.

The subproblem corresponding to  block $\left\{ {{y_{c,k}},{y_{r,n}},\forall k,\forall n} \right\}$ is given by 
\begin{subequations}\label{auxiliary_sub2}
	\begin{align}
	&\mathop {\max }\limits_{\left\{ {{y_{c,k}},{y_{r,n}}} \right\}} \sum\limits_{k = 1}^K {{{\left| {{y_{c,k}}} \right|}^2}}  + \sum\limits_{n = 1}^N {{{\left| {{y_{r,n}}} \right|}^2}}  - \frac{1}{{2\rho }}\left( {\sum\limits_{k = 1}^K {{{\left| {{{\bf{g}}^H}{{\bf{w}}_{c,k}} - {y_{c,k}}} \right|}^2}} } \right.\notag\\
		& \qquad\qquad \left. { + \sum\nolimits_{n = 1}^N {{{\left| {{{\bf{g}}^H}{{\bf{w}}_{r,n}} - {y_{r,n}}} \right|}^2}} } \right)\label{auxiliary_sub2_obj}\\
		&{\rm s.t.}~\eqref{type_I_P_new_const2}.
	\end{align}
\end{subequations}
It is observed that the objective function \eqref{auxiliary_sub2_obj} is a difference of two convex (DC) functions, which is non-convex.
 To solve it,  the successive convex approximation  (SCA) technique is applied.
Specifically, for any given points $y_{c,k}^r$ and $y_{r,n}^r$, we have 
\begin{align}
	&{\left| {{y_{c,k}}} \right|^2} \ge  - {\left| {y_{c,k}^r} \right|^2} + 2{\mathop{\rm Re}\nolimits} \left\{ {y_{c,k}^Hy_{c,k}^r} \right\} \overset{\triangle}{=} f_1^{{\rm{lb}}}\left( {{y_{c,k}}} \right), \forall k, \label{lower_bound1}\\
	&{\left| {{y_{r,n}}} \right|^2} \ge  - {\left| {y_{r,n}^r} \right|^2} + 2{\mathop{\rm Re}\nolimits} \left\{ {y_{r,n}^Hy_{r,n}^r} \right\} \overset{\triangle}{=} f_2^{{\rm{lb}}}\left( {{y_{r,n}}} \right),\forall n.\label{lower_bound2}
\end{align}  
As a result, problem \eqref{auxiliary_sub2} can be approximated as 
\begin{subequations}\label{auxiliary_sub2_lowbound}
	\begin{align}
	&\mathop {\max }\limits_{\left\{ {{y_{c,k}},{y_{r,n}}} \right\}} \sum\nolimits_{k = 1}^K {f_1^{{\rm{lb}}}\left( {{y_{c,k}}} \right)}  + \sum\nolimits_{n = 1}^N {f_2^{{\rm{lb}}}\left( {{y_{r,n}}} \right)}  - \frac{1}{{2\rho }}\times\notag\\
		& {\sum\nolimits_{k = 1}^K {{{\left| {{{\bf{g}}^H}{{\bf{w}}_{c,k}} - {y_{c,k}}} \right|}^2} + \sum\nolimits_{n = 1}^N {{{\left| {{{\bf{g}}^H}{{\bf{w}}_{r,n}} - {y_{r,n}}} \right|}^2}} } }\label{auxiliary_sub2_lowbound_obj}\\
		&\!\!\!\!{\rm s.t.}{\kern 1pt}{\left| {{y_{c,k}}} \right|^2} \le {r_{e,k,{\rm{th}}}}\left( {\sum\limits_{i \ne k}^K {f_1^{{\rm{lb}}}\left( {{y_{c,i}}} \right)} \! +\! \sum\limits_{n = 1}^N {f_2^{{\rm{lb}}}\left( {{y_{r,n}}} \right)}  \!+\! \sigma _t^2} \right),\forall k .
	\end{align}
\end{subequations}
It can be readily seen that problem \eqref{auxiliary_sub2_lowbound} is  a QCQP, which can be optimally  solved by the  interior-point method \cite{boyd2004convex}.

\subsubsection{Optimizing $\left\{ {{{\bf{w}}_{c,k}},{{\bf{w}}_{r,n}}} \right\}$ for given ${\left\{ {{v_m}} \right\}}$ and $\Omega $} This subproblem is
given by (dropping irrelevant constants w.r.t. $\left\{ {{{\bf{w}}_{c,k}},{{\bf{w}}_{r,n}}} \right\}$)
\begin{subequations} \label{Perfect_subproblem_1}
	\begin{align}
	&\mathop {\min }\limits_{\left\{ {{{\bf{w}}_{c,k}}} \right\},\left\{ {{{\bf{w}}_{r,n}}} \right\}} \sum\limits_{k = 1}^K {{{\left| {{{\bf{g}}^H}{{\bf{w}}_{c,k}} - {y_{c,k}}} \right|}^2} + } \sum\limits_{n = 1}^N {{{\left| {{{\bf{g}}^H}{{\bf{w}}_{r,n}} - {y_{r,n}}} \right|}^2} + }  \notag\\
			&\!\!\sum\limits_{k = 1}^K {\sum\limits_{i = 1}^K {{{\left| {{\bf{h}}_k^H{{\bf{w}}_{c,i}}\! - \!{z_{c,k,i}}} \right|}^2}} } + \sum\limits_{k = 1}^K {\sum\limits_{n = 1}^N {{{\left| {{\bf{h}}_k^H{{\bf{w}}_{r,n}} \!- \!{z_{r,k,n}}} \right|}^2}} }  \\
	&{\rm s.t.}~\eqref{type_I_P_const3}.
	\end{align}
\end{subequations}
Note that  problem \eqref{Perfect_subproblem_1} is also a QCQP, which  can be solved
by the interior point method but with  a high computational complexity  \cite{boyd2004convex}. To reduce the computational complexity, we obtain
a semi-closed-form yet optimal solution for the transmit
beamformers  by using the Lagrange duality method. By introducing the dual variable $\mu_2\ge0$ associated
with constraint \eqref{type_I_P_const3}, the Lagrangian function of problem \eqref{Perfect_subproblem_1} is given by
\begin{align}
&\!\!\!{\cal L}_2\left( {{{\bf{w}}_{c,k}},{{\bf{w}}_{r,n}},{\mu _2}} \right) \!= \!\sum\limits_{k = 1}^K {{{\left| {{{\bf{g}}^H}{{\bf{w}}_{c,k}}\! -\! {y_{c,k}}} \right|}^2}\! +\! \sum\limits_{n = 1}^N {{{\left| {{{\bf{g}}^H}{{\bf{w}}_{r,n}}\! -\! {y_{r,n}}} \right|}^2}} }   \notag\\
& + \sum\limits_{k = 1}^K {\sum\limits_{i = 1}^K {{{\left| {{\bf{h}}_k^H{{\bf{w}}_{c,i}} - {z_{c,k,i}}} \right|}^2}} }+ \sum\limits_{k = 1}^K {\sum\limits_{n = 1}^N {{{\left| {{\bf{h}}_k^H{{\bf{w}}_{r,n}} - {z_{r,k,n}}} \right|}^2}} }  + \notag\\
&{\mu _2}\left( {\sum\limits_{k = 1}^K {{{\left\| {{{\bf{w}}_{c,k}}} \right\|}^2}}  + \sum\limits_{n = 1}^N {{{\left\| {{{\bf{w}}_{r,n}}} \right\|}^2}}  - {P_{\max }}} \right).
\end{align}
By taking the first-order derivative of ${\cal L}_2\left( {{{\bf{w}}_{c,k}},{{\bf{w}}_{r,n}},{\mu _2}} \right)$  w.r.t. ${{{\bf{w}}_{c,k}}}$ and ${{{\bf{w}}_{r,n}}}$ and  setting both to zero, we obtain the optimal solutions as
\begin{align}
{\bf{w}}_{c,k}^{{\rm{opt}}}\left( {{\mu _2}} \right)& = {\left( {{\bf{g}}{{\bf{g}}^H} + \sum\nolimits_{i = 1}^K {{{\bf{h}}_i}{\bf{h}}_i^H + {\mu _2}{{\bf{I}}_N}} } \right)^{-1} }\notag\\
&\times\left( {{y_{c,k}}{\bf{g}} + \sum\nolimits_{i = 1}^K {{z_{c,i,k}}{{\bf{h}}_i}} } \right),  k \in {\cal K},\\
{\bf{w}}_{r,n}^{{\rm{opt}}}\left( {{\mu _2}} \right) &={\left( {{\bf{g}}{{\bf{g}}^H} + \sum\nolimits_{i = 1}^K {{{\bf{h}}_i}{\bf{h}}_i^H + {\mu _2}{{\bf{I}}_N}} } \right)^{-1} }\notag\\
&\times\left( {{y_{r,n}}{\bf{g}} + \sum\nolimits_{i = 1}^K {{z_{r,i,n}}{{\bf{h}}_i}} } \right),n \in {\cal N}.
\end{align}
Note that  the optimal solution must be satisfied with the following complementary slackness condition \cite{boyd2004convex}
\begin{align}
\mu _2^{{\rm{opt}}}\left( {P\left( {\mu _2^{{\rm{opt}}}} \right) - {P_{\max }}} \right) = 0,
\end{align}
where $P\left( {\mu _1^{{\rm{opt}}}} \right) = \sum\limits_{k = 1}^K {{{\left\| {{\bf{w}}_{c,k}^{{\rm{opt}}}\left( {\mu _2^{{\rm{opt}}}} \right)} \right\|}^2}}  + \sum\limits_{n = 1}^N {{{\left\| {{\bf{w}}_{r,n}^{{\rm{opt}}}\left( {\mu _2^{{\rm{opt}}}} \right)} \right\|}^2}} $. We first check whether $\mu _2^{{\rm{opt}}}=0$ is the optimal solution or not. If $P\left( 0 \right) - {P_{\max }}<0$, it means that the optimal dual variable $\mu _2^{{\rm{opt}}}$ equals $0$;  otherwise, the optimal $\mu _2^{{\rm{opt}}}$ is a positive value that satisfies $P\left( {\mu _2^{{\rm{opt}}}} \right) - {P_{\max }} = 0$, and can be obtained as follows.  Let ${\bf{S}} = {\bf{g}}{{\bf{g}}^H} + \sum\limits_{i = 1}^K {{{\bf{h}}_i}{\bf{h}}_i^H} $,  ${{\bf{t}}_{c,k}} = {y_{c,k}}{\bf{g}} + \sum\limits_{i = 1}^K {{z_{c,i,k}}{{\bf{h}}_i}} $, and ${{\bf{t}}_{r,n}} = {y_{r,n}}{\bf{g}} + \sum\limits_{i = 1}^K {{z_{r,i,n}}{{\bf{h}}_i}} $, which implies
\begin{align}
&{\left\| {{{\bf{w}}_{c,k}^{\rm opt}}\left( {{\mu _2}} \right)} \right\|^2} = {\rm{tr}}\left( {{{\left( {{\bf{S}} + {\mu _2}{{\bf{I}}_N}} \right)}^{ - 2}}{{\bf{t}}_{c,k}}{\bf{t}}_{c,k}^H} \right), \notag\\
&{\left\| {{{\bf{w}}_{r,n}^{\rm opt}}\left( {{\mu _2}} \right)} \right\|^2} = {\rm{tr}}\left( {{{\left( {{\bf{S}} + {\mu _2}{{\bf{I}}_N}} \right)}^{ - 2}}{{\bf{t}}_{r,n}}{\bf{t}}_{r,n}^H} \right). \label{Perfect_transmit_power}
\end{align}
Since ${\bf{S}}$ is a positive semi-definite matrix,   its eigendecomposition can be expressed as 
 ${\bf{S}} = {\bf{U\Sigma }}{{\bf{U}}^H}$. Substituting it into \eqref{Perfect_transmit_power}   yields 
%we  can perform eigenvalue decomposition on it as ${\bf{S}} = {\bf{U\Sigma }}{{\bf{U}}^H}$. Then, substituting ${\bf{S}} = {\bf{U\Sigma }}{{\bf{U}}^H}$ into \eqref{Perfect_transmit_power}, we have 
\begin{align}
\!\!\!\!\!P\left( {{\mu _2}} \right) = \sum\limits_{i = 1}^N {\frac{{{{\left( {{{\bf{U}}^H}\left( {\sum\limits_{k = 1}^K {{{\bf{t}}_{c,k}}{\bf{t}}_{c,k}^H}  + \sum\limits_{n = 1}^N {{{\bf{t}}_{r,n}}{\bf{t}}_{r,n}^H} } \right){\bf{U}}} \right)}_{i,i}}}}{{{{\left( {{{\bf{\Sigma }}_{i,i}} + {\mu _2}} \right)}^2}}}} . \label{upperbound_mu2}
\end{align}
It can be readily seen that $P\left( {{\mu _2}} \right)$ is monotonically decreasing w.r.t.  $\mu_2$, which motivates us to apply the bisection method to search for $\mu_2$  satisfying $P\left( {\mu _2^{{\rm{opt}}}} \right) = {P_{\max }}$. To reduce the search space, an upper bound of $\mu_2$ can be derived as $\mu _2^{{\rm{up}}} = \sqrt {\sum\limits_{i = 1}^N {{{\left( {{\bf{U}}^H\left( {\sum\limits_{k = 1}^K {{{\bf{t}}_{c,k}}{\bf{t}}_{c,k}^H}  + \sum\limits_{n = 1}^N {{{\bf{t}}_{r,n}}{\bf{t}}_{r,n}^H} } \right){{\bf{U}}}} \right)}_{i,i}}/{P_{\max }}} } $ by setting  ${{{\bf{\Sigma }}_{i,i}}}$ in  \eqref{upperbound_mu2} to zero. 
\subsubsection{Optimizing ${\left\{ {{v_m}} \right\}}$ for given $\left\{ {{{\bf{w}}_{c,k}},{{\bf{w}}_{r,n}}} \right\}$  and $\Omega $} This subproblem is
given by (ignoring the constant terms w.r.t. ${\left\{ {{v_m}} \right\}}$)
\begin{subequations}\label{Perfect_subproblem_2}
	\begin{align}
	&\mathop {\min }\limits_{ {\left\{ {{v_m}} \right\}}} \sum\limits_{k = 1}^K {{{\left| {{{\bf{g}}^H}{{\bf{w}}_{c,k}} - {y_{c,k}}} \right|}^2} + } \sum\limits_{n = 1}^N {{{\left| {{{\bf{g}}^H}{{\bf{w}}_{r,n}} - {y_{r,n}}} \right|}^2} + }  \notag\\
	&\!\!\!\sum\limits_{k = 1}^K {\sum\limits_{i = 1}^K {{{\left| {{\bf{h}}_k^H{{\bf{w}}_{c,i}} \!- \!{z_{c,k,i}}} \right|}^2}} } \!+\! \sum\limits_{k = 1}^K {\sum\limits_{n = 1}^N {{{\left| {{\bf{h}}_k^H{{\bf{w}}_{r,n}} \!- \!{z_{r,k,n}}} \right|}^2}} } \label{Perfect_subproblem_2_obj} \\
	&{\rm s.t.}~\eqref{type_I_P_const4}.
	\end{align}
\end{subequations}
Although the objective function \eqref{Perfect_subproblem_2_obj} is a  quadratic function of ${\bf v}$, the unit-modulus constraint imposed on each IRS phase shift in \eqref{type_I_P_const4} is non-convex.  Here, we construct  an upper-bounded convex surrogate function for 
\eqref{Perfect_subproblem_2_obj} by applying the MM algorithm \cite{7547360}, based on which  a closed-form solution for the IRS phase shifts is derived. Specifically, the surrogate
function at any given point ${{{\bf{v}}^r}}$, denoted by  $\varpi \left( {{\bf{v}}|{{\bf{v}}^r}} \right)$,  for a quadratic function ${{\bf{v}}^H}{\bf{Av}}$  can be expressed as
\begin{align}
	\varpi \left( {{\bf{v}}|{{\bf{v}}^r}} \right) &= {\lambda _{\max }}{{\bf{v}}^H}{\bf{v}} - 2{\mathop{\rm Re}\nolimits} \left\{ {{{\bf{v}}^H}\left( {{\lambda _{\max }}{\bf{I}}_M - {\bf{A}}} \right){{\bf{v}}^r}} \right\} \notag\\
	&+ {{\bf{v}}^{r,H}}\left( {{\lambda _{\max }}{\bf{I}}_M - {\bf{A}}} \right){{\bf{v}}^r},
\end{align}
where ${\bf{A}} \in {{\mathbb C}^{M \times M}}$ is   positive semi-definite, and ${\lambda _{\max }}$ is the maximum eigenvalue of ${\bf{A}}$. As a result, based on ${{\bf{v}}^H}{\bf{v}} = M$, we can solve the following approximate optimization problem (ignoring constant terms w.r.t. ${\left\{ {{v_m}} \right\}}$)
\begin{subequations}\label{Perfect_subproblem_2_upper}
\begin{align} 
&\mathop {\max }\limits_{{v_m}} {\mathop{\rm Re}\nolimits} \left\{ {{{\bf{v}}^H}{\bf{q}}^r} \right\}\\
&{\rm s.t.}~\eqref{type_I_P_const4},
\end{align}
\end{subequations}
where ${{\bf{q}}^r} = \sum\limits_{k = 1}^K {\left( {\left( {{\lambda _{\max ,1,k}}{{\bf{I}}_M} - {\Upsilon _{r,c,k}}} \right){{\bf{v}}^r} + y_{c,k}^H{{\bf{F}}_r}{{\bf{w}}_{c,k}}} \right)} $  

$ {\kern 1pt} {\kern 1pt} {\kern 1pt} {\kern 1pt}{\kern 1pt} {\kern 1pt} {\kern 1pt} {\kern 1pt} {\kern 1pt} {\kern 1pt} {\kern 1pt} {\kern 1pt}{\kern 1pt} {\kern 1pt} {\kern 1pt} {\kern 1pt} {\kern 1pt} {\kern 1pt} {\kern 1pt} {\kern 1pt}{\kern 1pt} {\kern 1pt} {\kern 1pt} {\kern 1pt} {\kern 1pt} {\kern 1pt} {\kern 1pt} {\kern 1pt} $
$ + \sum\limits_{n = 1}^N {\left( {\left( {{\lambda _{\max ,2,n}}{{\bf{I}}_M} - {\Upsilon _{r,r,n}}} \right){{\bf{v}}^r} + y_{r,n}^H{{\bf{F}}_r}{{\bf{w}}_{r,n}}} \right)} $ 

${\kern 1pt} {\kern 1pt} {\kern 1pt} {\kern 1pt} {\kern 1pt} {\kern 1pt} {\kern 1pt} {\kern 1pt}{\kern 1pt} {\kern 1pt} {\kern 1pt} {\kern 1pt} {\kern 1pt} {\kern 1pt} {\kern 1pt} {\kern 1pt}{\kern 1pt} {\kern 1pt} {\kern 1pt} {\kern 1pt} {\kern 1pt} {\kern 1pt} {\kern 1pt} {\kern 1pt}{\kern 1pt} {\kern 1pt} {\kern 1pt} {\kern 1pt} {\kern 1pt} {\kern 1pt} {\kern 1pt} {\kern 1pt} $$ + \sum\limits_{k = 1}^K {\sum\limits_{i = 1}^K {\left( {\left( {{\lambda _{\max ,3,k}}{{\bf{I}}_M} - {\Upsilon _{c,k,i}}} \right){{\bf{v}}^r}{\rm{ - }}{\Psi _{c,k,i}}} \right)} } $ 

${\kern 1pt} {\kern 1pt} {\kern 1pt} {\kern 1pt} {\kern 1pt} {\kern 1pt} {\kern 1pt} {\kern 1pt}{\kern 1pt} {\kern 1pt} {\kern 1pt} {\kern 1pt} {\kern 1pt} {\kern 1pt} {\kern 1pt} {\kern 1pt}{\kern 1pt} {\kern 1pt} {\kern 1pt} {\kern 1pt} {\kern 1pt} {\kern 1pt} {\kern 1pt} {\kern 1pt}{\kern 1pt} {\kern 1pt} {\kern 1pt} {\kern 1pt} {\kern 1pt} {\kern 1pt} {\kern 1pt} {\kern 1pt} $$ + \sum\nolimits_{k = 1}^K {\sum\nolimits_{n = 1}^N {\left( {\left( {{\lambda _{\max ,4,k,n}}{{\bf{I}}_M} \!-\! {\Upsilon _{r,k,n}}} \right){{\bf{v}}^r}\!-\!{\Psi _{r,n,k}}} \right)} } $,
${\Psi _{c,k,i}} = {{\bf{F}}_k}{{\bf{w}}_{c,i}}\left( {{\bf{w}}_{c,i}^H{{\bf{h}}_{d,k}} - z_{c,k,i}^H} \right)$, ${\Psi _{r,n,k}} = {{\bf{F}}_k}{{\bf{w}}_{r,n}}\left( {{\bf{w}}_{r,n}^H{{\bf{h}}_{d,k}} - z_{r,k,n}^H} \right)$, ${\Upsilon _{r,c,k}} = {{\bf{F}}_r}{{\bf{w}}_{c,k}}{\bf{w}}_{c,k}^H{\bf{F}}_r^H$, ${\Upsilon _{r,r,n}} = {{\bf{F}}_r}{{\bf{w}}_{r,n}}{\bf{w}}_{r,n}^H{\bf{F}}_r^H$, ${\Upsilon _{c,k,i}} = {{\bf{F}}_k}{{\bf{w}}_{c,i}}{\bf{w}}_{c,i}^H{\bf{F}}_k^H$, ${\Upsilon _{r,k,n}} = {{\bf{F}}_k}{{\bf{w}}_{r,n}}{\bf{w}}_{r,n}^H{\bf{F}}_k^H$,
 and ${{\lambda _{\max ,1,k}}}$, ${{\lambda _{\max ,2,n}}}$, ${{\lambda _{\max ,3,k,i}}}$, and ${{\lambda _{\max ,4,k,n}}}$ represent the  maximum eigenvalues of ${\Upsilon _{r,c,k}}$, ${\Upsilon _{r,r,n}}$, ${\Upsilon _{c,k,i}}$, and ${\Upsilon _{r,k,n}}$, respectively. The
optimal solution $\bf v$ to problem \eqref{Perfect_subproblem_2_upper} is then given by 
\begin{align}
 {{\bf{v}}^{{\rm{opt}}}} = {e^{  j\arg \left( {{{\bf{q}}^r}} \right)}}. \label{optimal_phase}
\end{align}
\subsection{Outer Layer Optimization}
In the outer layer, the penalty  parameter  in the $r$th iteration is   updated as follows
\begin{align}
	{\rho ^r} = c{\rho ^{r - 1}},~0 < c < 1, \label{update_penalty_coefficient}
\end{align}
where $c$ is a constant scaling factor that is used to control the convergence behavior.
\subsection{Overall Algorithm and Computational Complexity}
\begin{algorithm}[!t]
	\caption{Penalty-based algorithm for solving problem \eqref{type_I_P}.}
	\label{alg1}
	\begin{algorithmic}[1]
		\STATE  \textbf{Initialize} ${\bf v}$, $\left\{ {{{\bf{w}}_{c,k}},{{\bf{w}}_{r,n}}} \right\}$, $\left\{ {y_{c,k},y_{r,n}} \right\}$, $c$, $\rho$, $\varepsilon_{\rm in}$, and  $\varepsilon_{\rm out}$.
		\STATE  \textbf{repeat: outer layer}
		\STATE \quad \textbf{repeat: inner layer }
		\STATE  \qquad Update   auxiliary variables $\left\{ {{z_{c,k,i}},{z_{r,k,n}}} \right\}$  by solving \\
		\qquad problem \eqref{auxiliary_sub1}.
		\STATE  \qquad  Update   auxiliary variables $\left\{ {y_{c,k},y_{r,n}} \right\}$  by  solving\\
		\qquad   problem \eqref{auxiliary_sub2_lowbound}.
		\STATE  \qquad Update transmit beamformers $\left\{ {{{\bf{w}}_{c,k}},{{\bf{w}}_{r,n}}} \right\}$ by solving \\ \qquad problem  \eqref{Perfect_subproblem_1}.
		\STATE  \qquad Update IRS phase shifts $\{v_m\}$ based on \eqref{optimal_phase}.
		\STATE \quad \textbf{until} the fractional increase of the objective value of  \eqref{type_I_P_new_penbalty} \\
		\quad is below a threshold $\varepsilon_{\rm in}$.
		\STATE  \quad Update penalty parameter $\rho$ based on \eqref{update_penalty_coefficient}.
		\STATE \textbf{until}  termination indicator $\xi$ defined in   \eqref{termination_indicator} is below a threshold $\varepsilon_{\rm out}$.
	\end{algorithmic}
\end{algorithm}
The termination indicator for the penalty-based algorithm  is given  by 
\begin{align}
\xi  =& \mathop {\max }\limits_{\forall i,k,n} \left\{ {{{\left| {{{\bf{g}}^H}{{\bf{w}}_{c,k}} - {y_{c,k}}} \right|}^2},{{\left| {{{\bf{g}}^H}{{\bf{w}}_{r,n}} - {y_{r,n}}} \right|}^2},} \right.\notag\\
&\qquad~ \left. {{{\left| {{\bf{h}}_k^H{{\bf{w}}_{c,i}} - {z_{c,k,i}}} \right|}^2},{{\left| {{\bf{h}}_k^H{{\bf{w}}_{r,n}} - {z_{r,k,n}}} \right|}^2}} \right\}. \label{termination_indicator}
\end{align}
If $\xi$ is smaller than a predefined value,  constraint \eqref{type_I_P_new_const3} is considered to be met with equality for a given accuracy.
The  proposed  algorithm is  summarized in Algorithm~\ref{alg1}, whose computational complexity  is given by 
${\cal O}\left( {{I_{{\rm{out}}}}{I_{{\rm{in}}}}\left( {K{{\log }_2}\left( {\frac{1}{\varepsilon }} \right){N^3}{\rm{ + }}} \right.} \right.$ $\left. {\left. {{{\log }_2}\left( {\frac{{\mu _2^{{\rm{up}}}}}{\varepsilon }} \right){N^3} + {{\left( {K + N} \right)}^{3.5}} + {M^3}} \right)} \right)$, where $\varepsilon $ represents  the iteration accuracy, and  ${{I_{{\rm{in}}}}}$ and ${{I_{{\rm{out}}}}}$ denote the numbers of iterations required
for reaching convergence in the inner layer and outer layer, respectively.

\section{Proposed Solution for Imperfect CSI and  Uncertain Target Location}
 In this section, we consider   the case with imperfect CSI and   uncertain target location. Since problem  \eqref{Imperfect_P} involves an infinite number of inequalities in  constraints \eqref{Imperfect_const1} and \eqref{Imperfect_const2},
 the previous  penalty-based algorithm  is no longer   applicable   for solving problem  \eqref{Imperfect_P}, which thus calls for   new algorithm design.
By introducing auxiliary variables $\left\{ {{\beta _{c,k}}}\ge 0 \right\}$ satisfying ${\beta _{c,k}} = \sum\limits_{i \ne k}^K {{{\left| {{\bf{h}}_k^H{{\bf{w}}_{c,i}}} \right|}^2}}  + \sum\limits_{n = 1}^N {{{\left| {{\bf{h}}_k^H{{\bf{w}}_{r,n}}} \right|}^2}}  + \sigma _k^2,k\in{\cal K}$, constraint \eqref{Imperfect_const1} can be equivalently transformed as
 \begin{align}
 &\!\!\!\!\!{\left| {{\bf{h}}_k^H{{\bf{w}}_{c,k}}} \right|^2} \ge {\beta _{c,k}}{r_{k,{\rm{th}}}}, \Delta {{\bf{h}}_{d,k}} \in  {{\bf{\cal H}}_{d,k}}, \Delta {{\bf{F}}_k} \in {{\bf{\cal F}}_k}, k\in{\cal K},\label{type_I_P_const1_new1}\\
& \!\!\!\!\!\sum\limits_{i \ne k}^K {{{\left| {{\bf{h}}_k^H{{\bf{w}}_{c,i}}} \right|}^2}}  + \sum\limits_{n = 1}^N {{{\left| {{\bf{h}}_k^H{{\bf{w}}_{r,n}}} \right|}^2}}  + \sigma _k^2 \le {\beta _{c,k}}, \Delta {{\bf{h}}_{d,k}} \in  {{\bf{\cal H}}_{d,k}},\notag\\
&\qquad \qquad\qquad\qquad\qquad\qquad\qquad\quad \Delta {{\bf{F}}_k} \in {{\bf{\cal F}}_k}, k\in{\cal K}.\label{type_I_P_const1_new2}
 \end{align}
 Although the left-hand side of  \eqref{type_I_P_const1_new1} is convex w.r.t $\bf v$ (recall that ${\bf{h}}_k^H = {{\bf{v}}^H}{{\bf{F}}_k} + {\bf{h}}_{d,k}^H$), the resulting set is not a convex set since the superlevel
 set of a convex quadratic function is not convex in general. To address this  non-convex constraint, we take
  the first-order Taylor expansion of  ${\left| {{\bf{h}}_k^H{{\bf{w}}_{c,k}}} \right|^2}$  at any given feasible point ${{{\bf{v}}^{r}}}$ to obtain the following lower bound 
 \begin{align}
 &\!\!\!\!\!{\left| {{\bf{h}}_k^H{{\bf{w}}_{c,k}}} \right|^2} \ge  f_k^{{\rm{lb}}}\left( {\bf{v}} \right)\overset{\triangle}{=} - {\left| {{{\bf{v}}^{r,H}}{{\bf{F}}_k}{{\bf{w}}_{c,k}^r} + {\bf{h}}_{d,k}^H{{\bf{w}}_{c,k}^r}} \right|^2}+2{\mathop{\rm Re}\nolimits}\Big\{ \notag\\
 &\!\!\!{{{\left( {{{\bf{v}}^H}{{\bf{F}}_k}{{\bf{w}}_{c,k}} + {\bf{h}}_{d,k}^H{{\bf{w}}_{c,k}}} \right)}^H}\left( {{{\bf{v}}^{r,H}}{{\bf{F}}_k}{{\bf{w}}_{c,k}^r} + {\bf{h}}_{d,k}^H{{\bf{w}}_{c,k}^r}} \right)} \Big\} \label{lowerbound_0}
 \end{align}
 which is linear and convex w.r.t. ${{{\bf{v}}}}$.
 
Substituting  ${{\bf{F}}_k}{\rm{ = }}{{{\bf{\hat F}}}_k} + \Delta {{\bf{F}}_k}$ and ${{\bf{h}}_{d,k}} = {{{\bf{\hat h}}}_{d,k}} + \Delta {{\bf{h}}_{d,k}}$ into term ${\left| {{{\bf{v}}^{r,H}}{{\bf{F}}_k}{{\bf{w}}_{c,k}^r} + {\bf{h}}_{d,k}^H{{\bf{w}}_{c,k}^r}} \right|^2}$ in \eqref{lowerbound_0},  we can rewrite it  as 
  \begin{align}
&{\left| {{{\bf{v}}^{r,H}}{{\bf{F}}_k}{{\bf{w}}_{c,k}^r} + {\bf{h}}_{d,k}^H{{\bf{w}}_{c,k}^r}} \right|^2} = {\left| {{\bf{\hat h}}_k^{r,H}{{\bf{w}}_{c,k}^r}} \right|^2} +\notag\\
&   2{\rm{Re}}{\left\{ {{{\left( {{\bf{\hat h}}_k^{r,H}{{\bf{w}}_{c,k}^r}} \right)}^H}\left( {{{\bf{v}}^{r,H}}\Delta {{\bf{F}}_k}{{\bf{w}}_{c,k}^r} + \Delta {\bf{h}}_{d,k}^H{{\bf{w}}_{c,k}^r}} \right)} \right\}} + \notag\\
&{\left| {{{\bf{v}}^{r,H}}\Delta {{\bf{F}}_k}{{\bf{w}}_{c,k}^r} + \Delta {\bf{h}}_{d,k}^H{{\bf{w}}_{c,k}^r}} \right|^2}, \label{expression_0}
  \end{align}
 where ${\bf{\hat h}}_k^{r,H} = {{\bf{v}}^{r,H}}{{{\bf{\hat F}}}_k} + {\bf{\hat h}}_{d,k}^H$. Below, we rewrite  terms in  \eqref{expression_0} into 
  a compact form that facilitates the  algorithm design. Specifically, we first expand ${\left| {{{\bf{v}}^{r,H}}\Delta {{\bf{F}}_k}{{\bf{w}}_{c,k}^r} + \Delta {\bf{h}}_{d,k}^H{{\bf{w}}_{c,k}^r}} \right|^2}$ as 
\begin{align}
&\!\!\!\!{\left| {{{\bf{v}}^{r,H}}\Delta {{\bf{F}}_k}{{\bf{w}}_{c,k}^r} + \Delta {\bf{h}}_{d,k}^H{{\bf{w}}_{c,k}^r}} \right|^2} = {{\bf{v}}^{r,H}}\Delta {{\bf{F}}_k}{{\bf{w}}_{c,k}^r}{\bf{w}}_{c,k}^{r,H}\Delta {\bf{F}}_k^H{{\bf{v}}^r} + \notag \\
&\Delta{\bf{h}}_{d,k}^H{{\bf{w}}_{c,k}^r}{\bf{w}}_{c,k}^{r,H}\Delta {{\bf{h}}_{d,k}}+ {{\bf{v}}^{r,H}}\Delta {{\bf{F}}_k}{{\bf{w}}_{c,k}^r}{\bf{w}}_{c,k}^{r,H}\Delta {{\bf{h}}_{d,k}} + \notag\\
&\Delta {\bf{h}}_{d,k}^H{{\bf{w}}_{c,k}^r}{\bf{w}}_{c,k}^{r,H}\Delta {\bf{F}}_k^H{{\bf{v}}^r},
\end{align}
where
\begin{align}
&{{\bf{v}}^{r,H}}\Delta {{\bf{F}}_k}{\bf{w}}_{c,k}^r{\bf{w}}_{c,k}^{r,H}\Delta {\bf{F}}_k^H{{\bf{v}}^r} = \notag\\
&\quad {\rm{ve}}{{\rm{c}}^H}\left( {\Delta {\bf{F}}_k^*} \right)\left( {{\bf{w}}_{c,k}^r{\bf{w}}_{c,k}^{r,H}} \right) \otimes \left( {{{\bf{v}}^{r,*}}{{\bf{v}}^{r,T}}} \right){\rm{vec}}\left( {\Delta {\bf{F}}_k^*} \right),\\
&{{\bf{v}}^{r,H}}\Delta {{\bf{F}}_k}{\bf{w}}_{c,k}^r{\bf{w}}_{c,k}^{r,H}\Delta {{\bf{h}}_{d,k}}=\notag\\
&\qquad{{\rm{vec}}^H}\left( {\Delta {\bf{F}}_k^*} \right)\left( {\left( {{\bf{w}}_{c,k}^r{\bf{w}}_{c,k}^{r,H}} \right) \otimes {{\bf{v}}^{r,*}}} \right)\Delta {{\bf{h}}_{d,k}},\\
& \Delta {\bf{h}}_{d,k}^H{\bf{w}}_{c,k}^r{\bf{w}}_{c,k}^{r,H}\Delta {\bf{F}}_k^H{{\bf{v}}^r} =\notag\\
&\qquad  \Delta {\bf{h}}_{d,k}^H\left( {\left( {{\bf{w}}_{c,k}^r{\bf{w}}_{c,k}^{r,H}} \right) \otimes {{\bf{v}}^{r,T}}} \right){\rm{vec}}\left( {\Delta {\bf{F}}_k^*} \right).
\end{align}
Thus, we can rewrite ${\left| {{{\bf{v}}^{r,H}}\Delta {{\bf{F}}_k}{{\bf{w}}_{c,k}^r} + \Delta {\bf{h}}_{d,k}^H{{\bf{w}}_{c,k}^r}} \right|^2}$ in a more compact form given by 
\begin{align}
{\left| {{{\bf{v}}^{r,H}}\Delta {{\bf{F}}_k}{{\bf{w}}_{c,k}^r} + \Delta {\bf{h}}_{d,k}^H{{\bf{w}}_{c,k}^r}} \right|^2} = \Delta {\bf{h}}_{k,{\rm{eff}}}^H{{\bf{H}}_{c,k}^r}\Delta {{\bf{h}}_{k,{\rm{eff}}}}, \label{expression_1}
\end{align}
where $\Delta {\bf{h}}_{k,{\rm{eff}}}^H = \left[ {\Delta {\bf{h}}_{d,k}^H{\kern 1pt} {\kern 1pt} {\kern 1pt} {\kern 1pt} {\kern 1pt} {\rm{ve}}{{\rm{c}}^H}\left( {\Delta {\bf{F}}_k^*} \right){\kern 1pt} } \right]$,  ${\bf{H}}_{c,k}^r = \left[ {\begin{array}{*{20}{c}}
	{{\bf{w}}_{c,k}^r{\bf{w}}_{c,k}^{r,H}}&{\left( {{\bf{w}}_{c,k}^r{\bf{w}}_{c,k}^{r,H}} \right) \otimes {{\bf{v}}^{r,T}}}\\
	{\left( {{\bf{w}}_{c,k}^r{\bf{w}}_{c,k}^{r,H}} \right) \otimes {{\bf{v}}^{r,*}}}&{\left( {{\bf{w}}_{c,k}^r{\bf{w}}_{c,k}^{r,H}} \right) \otimes \left( {{{\bf{v}}^{r,{\rm{*}}}}{{\bf{v}}^{r,T}}} \right)}
	\end{array}} \right]$.
Then,   ${\left( {{\bf{\hat h}}_k^{r,H}{{\bf{w}}_{c,k}^r}} \right)^H}\left( {{{\bf{v}}^{r,H}}\Delta {{\bf{F}}_k}{{\bf{w}}_{c,k}^r} + \Delta {\bf{h}}_{d,k}^H{{\bf{w}}_{c,k}^r}} \right)$ can be expressed  as 
\begin{align}
&{\left( {{\bf{\hat h}}_k^{r,H}{{\bf{w}}_{c,k}^r}} \right)^H}\left( {{{\bf{v}}^{r,H}}\Delta {{\bf{F}}_k}{{\bf{w}}_{c,k}^r} + \Delta {\bf{h}}_{d,k}^H{{\bf{w}}_{c,k}^r}} \right) =\notag\\
& \qquad\qquad\qquad\qquad\qquad\quad {\left( {{\bf{\hat h}}_k^{r,H}{{\bf{w}}_{c,k}^r}} \right)^H}\Delta {\bf{h}}_{k,{\rm{eff}}}^H{{\bf{h}}_{c,k}^r}, \label{expression_2}
\end{align}
where ${{\bf{h}}_{c,k}^r} = {\left[ {{\bf{w}}_{c,k}^{r,T}{\kern 1pt} {\kern 1pt} {\kern 1pt} {\kern 1pt} \left( {{\bf{w}}_{c,k}^{r,T} \otimes {{\bf{v}}^{r,H}}} \right)} \right]^T}$.

Based on \eqref{expression_0}, \eqref{expression_1}, and \eqref{expression_2}, we can compactly rewrite ${\left| {{{\bf{v}}^{r,H}}{{\bf{F}}_k}{{\bf{w}}_{c,k}^r} + {\bf{h}}_{d,k}^H{{\bf{w}}_{c,k}^r}} \right|^2}$ in \eqref{expression_0} as 
\begin{align}
&{\left| {{{\bf{v}}^{r,H}}{{\bf{F}}_k}{{\bf{w}}_{c,k}^r} + {\bf{h}}_{d,k}^H{{\bf{w}}_{c,k}^r}} \right|^2} = \Delta {\bf{h}}_{k,{\rm{eff}}}^H{{\bf{H}}_{c,k}^r}\Delta {{\bf{h}}_{k,{\rm{eff}}}} +\notag\\ 
&\qquad  2{\rm{Re}}{\left\{ {{{\left( {{\bf{\hat h}}_k^{r,H}{{\bf{w}}_{c,k}^r}} \right)}^H}\Delta {\bf{h}}_{k,{\rm{eff}}}^H{{\bf{h}}_{c,k}^r}} \right\}}+{\left| {{\bf{\hat h}}_k^{r,H}{{\bf{w}}_{c,k}^r}} \right|^2} .\label{lowerboud_1}
\end{align}
In addition, we can expand ${\left( {{{\bf{v}}^H}{{\bf{F}}_k}{{\bf{w}}_{c,k}} + {\bf{h}}_{d,k}^H{{\bf{w}}_{c,k}}} \right)^H}$
\noindent$\times\left( {{{\bf{v}}^{r,H}}{{\bf{F}}_k}{{\bf{w}}_{c,k}^r} + {\bf{h}}_{d,k}^H{{\bf{w}}_{c,k}^r}} \right)$ in \eqref{lowerbound_0} as 
\begin{align}
&{\left( {{{\bf{v}}^H}{{\bf{F}}_k}{{\bf{w}}_{c,k}} + {\bf{h}}_{d,k}^H{{\bf{w}}_{c,k}}} \right)^H}\left( {{{\bf{v}}^{r,H}}{{\bf{F}}_k}{{\bf{w}}_{c,k}^r} + {\bf{h}}_{d,k}^H{{\bf{w}}_{c,k}^r}} \right) =\notag\\ &{\bf{w}}_{c,k}^H{{{\bf{\hat h}}}_k}{\bf{\hat h}}_k^{r,H}{{\bf{w}}_{c,k}^r} + {\bf{w}}_{c,k}^H{{{\bf{\hat h}}}_k}{{\bf{v}}^{r,H}}\Delta {{\bf{F}}_k}{{\bf{w}}_{c,k}^r}+ {\bf{w}}_{c,k}^H{{{\bf{\hat h}}}_k}\Delta {\bf{h}}_{d,k}^H{{\bf{w}}_{c,k}^r}\notag\\
&  + {\bf{w}}_{c,k}^H\Delta {\bf{F}}_k^H{\bf{v\hat h}}_k^{r,H}{{\bf{w}}_{c,k}^r} + {\bf{w}}_{c,k}^H\Delta {\bf{F}}_k^H{\bf{v}}{{\bf{v}}^{r,H}}\Delta {{\bf{F}}_k}{{\bf{w}}_{c,k}^r} + \notag\\
&{\bf{w}}_{c,k}^H\Delta {\bf{F}}_k^H{\bf{v}}\Delta {\bf{h}}_{d,k}^H{{\bf{w}}_{c,k}^r}+ {\bf{w}}_{c,k}^H\Delta {{\bf{h}}_{d,k}}{\bf{\hat h}}_k^{r,H}{{\bf{w}}_{c,k}^r} +  \notag\\
&{\bf{w}}_{c,k}^H\Delta {{\bf{h}}_{d,k}}{{\bf{v}}^{r,H}}\Delta {{\bf{F}}_k}{{\bf{w}}_{c,k}^r} +{\bf{w}}_{c,k}^H\Delta {{\bf{h}}_{d,k}}\Delta {\bf{h}}_{d,k}^H{{\bf{w}}_{c,k}^r}, \label{lowerbound_0_term1}
\end{align}
where ${\bf{\hat h}}_k^H = {{\bf{v}}^H}{{{\bf{\hat F}}}_k} + {\bf{\hat h}}_{d,k}^H$. Similarly, we can transform  terms in \eqref{lowerbound_0_term1} as 
\begin{align}
&\!\!{\bf{w}}_{c,k}^H{{{\bf{\hat h}}}_k}{{\bf{v}}^{r,H}}\Delta {{\bf{F}}_k}{{\bf{w}}_{c,k}^r} =
 {\rm{ve}}{{\rm{c}}^H}\left( {\Delta {\bf{F}}_k^ * } \right){\left( {{\bf{w}}_{c,k}^{r,T} \otimes {{\bf{v}}^{r,H}}} \right)^T}{\bf{w}}_{c,k}^H{{{\bf{\hat h}}}_k},\\
&\!\!{\bf{w}}_{c,k}^H\Delta {\bf{F}}_k^H{\bf{v\hat h}}_k^{r,H}{{\bf{w}}_{c,k}^r} = 
 {\bf{\hat h}}_k^{r,H}{{\bf{w}}_{c,k}^r}\left( {{\bf{w}}_{c,k}^H \otimes {{\bf{v}}^T}} \right){\rm{ve}}{{\rm{c}}}\left( {\Delta {{\bf{F}}_k^ *}} \right),\\
&\!\!{\bf{w}}_{c,k}^H\Delta {\bf{F}}_k^H{\bf{v}}{{\bf{v}}^{r,H}}\Delta {{\bf{F}}_k}{{\bf{w}}_{c,k}^r} = \notag\\
& \quad {\rm{ve}}{{\rm{c}}^H}\left( {\Delta {\bf{F}}_k^ * } \right){\left( {{{\left( {{{\bf{w}}_{c,k}^r}{\bf{w}}_{c,k}^H} \right)}^T} \otimes \left( {{\bf{v}}{{\bf{v}}^{r,H}}} \right)} \right)^T}{\rm{vec}}\left( {\Delta {\bf{F}}_k^ * } \right),\\
&{\bf{w}}_{c,k}^H\Delta {\bf{F}}_k^H{\bf{v}}\Delta {\bf{h}}_{d,k}^H{{\bf{w}}_{c,k}^r} = \notag\\
&\qquad\qquad\qquad~  \Delta {\bf{h}}_{d,k}^H{\left( {{{\left( {{{\bf{w}}_{c,k}^r}{\bf{w}}_{c,k}^H} \right)}^T} \otimes {\bf{v}}} \right)^T}{\rm{vec}}\left( {\Delta {\bf{F}}_k^*} \right).
\end{align}
Thus,  ${\left( {{{\bf{v}}^H}{{\bf{F}}_k}{{\bf{w}}_{c,k}} + {\bf{h}}_{d,k}^H{{\bf{w}}_{c,k}}} \right)^H}\left( {{{\bf{v}}^{r,H}}{{\bf{F}}_k}{{\bf{w}}_{c,k}^r} + {\bf{h}}_{d,k}^H{{\bf{w}}_{c,k}^r}} \right)$ can be written in a more compact form given by 
\begin{align}
&{\left( {{{\bf{v}}^H}{{\bf{F}}_k}{{\bf{w}}_{c,k}} + {\bf{h}}_{d,k}^H{{\bf{w}}_{c,k}}} \right)^H}\left( {{{\bf{v}}^{r,H}}{{\bf{F}}_k}{{\bf{w}}_{c,k}^r} + {\bf{h}}_{d,k}^H{{\bf{w}}_{c,k}^r}} \right) =  \notag\\
&\Delta {\bf{h}}_{k,{\rm{eff}}}^H{{\bf{H}}_{c,k}}\Delta {{\bf{h}}_{k,{\rm{eff}}}} +\Delta {\bf{h}}_{k,{\rm{eff}}}^H{{\bf{h}}_{c,k}^r}{\bf{w}}_{c,k}^H{{{\bf{\hat h}}}_k} +\notag\\
& {\bf{\hat h}}_k^{r,H}{{\bf{w}}_{c,k}^r}{\bf{h}}_{c,k}^H\Delta {{\bf{h}}_{k,{\rm{eff}}}} + {\bf{w}}_{c,k}^H{{{\bf{\hat h}}}_k}{\bf{\hat h}}_k^{r,H}{{\bf{w}}_{c,k}^r},\label{lowerboud_2}
\end{align}
where ${{\bf{h}}_{c,k}} = {\left[ {{\bf{w}}_{c,k}^{T}{\kern 1pt} {\kern 1pt} {\kern 1pt} {\kern 1pt} \left( {{\bf{w}}_{c,k}^{T} \otimes {{\bf{v}}^{H}}} \right)} \right]^T}$ and ${{\bf{H}}_{c,k}} = \left[ {\begin{array}{*{20}{c}}
	{{\bf{w}}_{c,k}^r{\bf{w}}_{c,k}^H}&{\left( {{\bf{w}}_{c,k}^r{\bf{w}}_{c,k}^H} \right) \otimes {{\bf{v}}^T}}\\
	{\left( {{\bf{w}}_{c,k}^r{\bf{w}}_{c,k}^H} \right) \otimes {{\bf{v}}^{r,*}}}&{\left( {{\bf{w}}_{c,k}^r{\bf{w}}_{c,k}^H} \right) \otimes \left( {{{\bf{v}}^{r,*}}{{\bf{v}}^T}} \right)}
	\end{array}} \right]$.

As a result, based on \eqref{lowerbound_0}, \eqref{lowerboud_1}, and \eqref{lowerboud_2},  constraint \eqref{type_I_P_const1_new1} can be approximated  as 
\begin{align}
&\Delta {\bf{h}}_{k,{\rm{eff}}}^H\left( {{{\bf{H}}_{c,k}} + {\bf{H}}_{c,k}^H - {\bf{H}}_{c,k}^r} \right)\Delta {{\bf{h}}_{k,{\rm{eff}}}}  + 2{\rm{Re}}\left\{ {{\bf{\hat h}}_{c,k}^H\Delta {{\bf{h}}_{k,{\rm{eff}}}}} \right\}   \notag\\
& + {{\bar h}_{c,k}} \ge {\beta _{c,k}}{r_{k,{\rm{th}}}}, \Delta {{\bf{h}}_{d,k}} \in  {{\bf{\cal H}}_{d,k}}, \Delta {{\bf{F}}_k} \in {{\bf{\cal F}}_k}, k\in{\cal K},\label{const1_1}
\end{align}
where ${\bf{\hat h}}_{c,k}^H = {\bf{\hat h}}_k^H{{\bf{w}}_{c,k}}{\bf{h}}_{c,k}^{r,H} + {\bf{\hat h}}_k^{r,H}{\bf{w}}_{c,k}^r{\bf{h}}_{c,k}^H - {\bf{\hat h}}_k^{r,H}{\bf{w}}_{c,k}^r{\bf{h}}_{c,k}^{r,H}$ and ${{\bar h}_{c,k}} = 2{\rm{Re}}\left\{ {{\bf{w}}_{c,k}^H{{{\bf{\hat h}}}_k}{\bf{\hat h}}_k^{r,H}{\bf{w}}_{c,k}^r} \right\} - {\left| {{\bf{\hat h}}_k^{r,H}{\bf{w}}_{c,k}^r} \right|^2}$.
We note that \eqref{const1_1}   still involves an infinite number of inequality constraints. To circumvent this difficulty, we convert the infinite number of constraints in  \eqref{const1_1} into an equivalent form with only a finite number of  LMIs  by applying
the following lemma.

\textbf{\emph{Lemma 1:}} (\textit{General} $\cal S$\textit{-Procedure}\cite{boyd1994linear}) Let ${f_i}\left( {\bf{z}} \right) = {{\bf{z}}^H}{{\bf{A}}_i}{\bf{z}} + 2{\mathop{\rm Re}\nolimits} \left\{ {{\bf{b}}_i^H{\bf{z}}} \right\} + {c_i},i \in \left\{ {0,1, \ldots ,{\rm{I}}} \right\}$, where ${\bf{z}} \in {{\mathbb C}^{N \times 1}}$ and  ${{\bf{A}}_i} = {\bf{A}}_i^H \in {{\mathbb C}^{N \times N}}$. The condition $\left\{ {{f_1}\left( {\bf{z}} \right) \ge 0} \right\}_{i = 1}^{{\rm{I}}} \Rightarrow {f_0}\left( {\bf{z}} \right) \ge 0$ holds if and only if there exist  ${\lambda _i} \ge 0,i \in \left\{ {1, \ldots ,{\rm{I}}} \right\}$ such that 
\begin{align}
\left[ {\begin{array}{*{20}{c}}
	{{{\bf{A}}_0}}&{{{\bf{b}}_0}}\\
	{{\bf{b}}_0^H}&{{c_0}}
	\end{array}} \right] - \sum\limits_{i = 1}^{{\rm{I}}} {{\lambda _i}\left[ {\begin{array}{*{20}{c}}
		{{{\bf{A}}_i}}&{{{\bf{b}}_i}}\\
		{{\bf{b}}_i^H}&{{c_i}}
		\end{array}} \right]}  \succeq {{\bf{0}}_{ {N + 1} }}.
\end{align}
Before applying Lemma 1,  we first   re-express  uncertainties   $\Delta {\bf{h}}_{d,k}\in  {\bf{\cal H}}_{d,k}$ and ${\Delta {\bf{F}}_k}\in { {\bf{\cal F}}_k}$  as 
\begin{align}
&\Delta {{\bf{h}}_{d,k}} \in  {{\bf{\cal H}}_{d,k}}  \Rightarrow\notag\\
& \Delta {\bf{h}}_{k,{\rm{eff}}}^H\left[ {\begin{array}{*{20}{c}}
	{{{\bf{I}}_N}}&{{{\bf{0}}_{N \times MN}}}\\
	{{{\bf{0}}_{MN \times N}}}&{{{\bf{0}}_{MN}}}
	\end{array}} \right]\Delta {{\bf{h}}_{k,{\rm{eff}}}} \le \varepsilon _{d,k}^2, k \in {\cal K},\label{const1_2}\\
&\Delta {{\bf{F}}_k} \in {{\bf{\cal F}}_k} \Rightarrow\notag\\
&  \Delta {\bf{h}}_{k,{\rm{eff}}}^H\left[ {\begin{array}{*{20}{c}}
	{{{\bf{0}}_N}}&{{{\bf{0}}_{N \times MN}}}\\
	{{{\bf{0}}_{MN \times N}}}&{{{\bf{I}}_{MN}}}
	\end{array}} \right]\Delta {{\bf{h}}_{k,{\rm{eff}}}} \le \varepsilon _k^2,k \in {\cal K}.\label{const1_3}
\end{align}
Then, based on Lemma 1, \eqref{const1_1} can be transformed as 
\begin{align}
&\left[ {\begin{array}{*{20}{c}}
	{{{\bf{H}}_{c,k}} + {\bf{H}}_{c,k}^H - {\bf{H}}_{c,k}^r + \left[ {\begin{array}{*{20}{c}}
			{{\lambda _{1,k}}{{\bf{I}}_N}}&{{{\bf{0}}_{N \times MN}}}\\
			{{{\bf{0}}_{MN \times N}}}&{{\lambda _{2,k}}{{\bf{I}}_{MN}}}
			\end{array}} \right]}&{{{{\bf{\hat h}}}_{c,k}}}\\
	{{\bf{\hat h}}_{c,k}^H}&{{c_k}}
	\end{array}} \right] \notag\\
&\qquad\qquad\qquad\qquad\qquad\qquad\quad \succeq {{\bf{0}}_{N + MN + 1}}, k \in {\cal K}, \label{const1_1_eqv}
\end{align}
where ${c_k} = {{\bar h}_{c,k}} - {\beta _{c,k}}{r_{k,{\rm{th}}}} - {\lambda _{1,k}}\varepsilon _{d,k}^2 - {\lambda _{2,k}}\varepsilon _k^2$,  $\lambda_{1,k}\ge0$ and $\lambda_{1,k}\ge0$ represents the auxiliary variables corresponding to \eqref{const1_2} and \eqref{const1_3}, respectively. It can be observed  that 
\eqref{const1_1_eqv} involves a finite number of LMIs, which thus can be handled using   convex optimization  techniques.

To tackle constraint \eqref{type_I_P_const1_new2}, we first equivalently  transform   it into LMIs based on the Schur’s complement given by 
\begin{align}
&\left[ {\begin{array}{*{20}{c}}
	{{\beta _{c,k}} - \sigma _k^2}&{{\bf{h}}_k^H{{\bf{W}}_{ - k}}}\\
	{{\bf{W}}_{ - k}^H{{\bf{h}}_k}}&{{{\bf{I}}_{K - 1 + N}}}
	\end{array}} \right] \succeq {{\bf{0}}_{K + N}}, \Delta {{\bf{h}}_{d,k}} \in  {{\bf{\cal H}}_{d,k}}, \notag\\
&\qquad\qquad\qquad\qquad\qquad\qquad\qquad\Delta {{\bf{F}}_k} \in {{\bf{\cal F}}_k}, k\in{\cal K}, \label{type_I_P_1}
\end{align}
where ${{\bf{W}}_{ - k}} = \left[ {{{\bf{w}}_{c,1}}, \ldots ,{{\bf{w}}_{c,k - 1}},{{\bf{w}}_{c,k + 1}}, \ldots ,{{\bf{w}}_{c,K}},{{\bf{w}}_{r,1}},} \right.$
\noindent$\left. { \ldots ,{{\bf{w}}_{r,N}}} \right]$.
Substituting  ${{\bf{F}}_k}{\rm{ = }}{{{\bf{\hat F}}}_k} + \Delta {{\bf{F}}_k}$ and ${{\bf{h}}_{d,k}} = {{{\bf{\hat h}}}_{d,k}} + \Delta {{\bf{h}}_{d,k}}$ into \eqref{type_I_P_1}, this can then be expanded as 
\begin{align}
 &\!\!\!\!\!\left[ {\begin{array}{*{20}{c}}
 	{{\beta _{c,k}} - \sigma _k^2}&{\left( {{{\bf{v}}^H}{{{\bf{\hat F}}}_k} + {\bf{\hat h}}_{d,k}^H} \right){{\bf{W}}_{ - k}}}\\
 	{{\bf{W}}_{ - k}^H\left( {{\bf{\hat F}}_k^H{\bf{v}} + {{{\bf{\hat h}}}_{d,k}}} \right)}&{{{\bf{I}}_{K - 1 + N}}}
 	\end{array}} \right] + \left[ {\begin{array}{*{20}{c}}
 	{{{\bf{0}}_{1 \times N}}}\\
 	{{\bf{W}}_{ - k}^H}
 	\end{array}} \right]\notag \\ 
&\!\!\!\!\! \times \Delta {{\bf{h}}_{d,k}}\left[ {\begin{array}{*{20}{c}}
	1&{{{\bf{0}}_{1 \times \left( {K - 1 + N} \right)}}}
	\end{array}} \right]+ \left[ {\begin{array}{*{20}{c}}
	1\\
	{{{\bf{0}}_{\left( {K - 1 + N} \right) \times 1}}}
	\end{array}} \right]\Delta {\bf{h}}_{d,k}^H\times
\notag\\
&\!\!\!\!\! \left[ {\begin{array}{*{20}{c}}
	{{{\bf{0}}_{N \times 1}}}&{{{\bf{W}}_{ - k}}}
	\end{array}} \right] + \left[ {\begin{array}{*{20}{c}}
	{{{\bf{0}}_{1 \times N}}}\\
	{{\bf{W}}_{ - k}^H}
	\end{array}} \right]\Delta {\bf{F}}_k^H\left[ {\begin{array}{*{20}{c}}
	{\bf{v}}&{{{\bf{0}}_{M \times \left( {K - 1 + N} \right)}}}
	\end{array}} \right]+ \notag\\
&\!\!\!\!\!\left[ {\begin{array}{*{20}{c}}
	{{{\bf{v}}^H}}\\
	{{{\bf{0}}_{\left( {K - 1 + N} \right) \times M}}}
	\end{array}} \right]\Delta {{\bf{F}}_k}\left[ {\begin{array}{*{20}{c}}
	{{{\bf{0}}_{N \times 1}}}&{{{\bf{W}}_{ - k}}}
	\end{array}} \right] \succeq  {{\bf{0}}_{K+N}}, \notag\\
&\qquad\qquad\qquad\qquad\quad \Delta {{\bf{h}}_{d,k}} \in  {{\bf{\cal H}}_{d,k}}, \Delta {{\bf{F}}_k} \in {{\bf{\cal F}}_k}, k\in{\cal K}. \label{type_I_P_1_expanded}
\end{align}
To address the infinite number of LMIs in \eqref{type_I_P_1_expanded}, we transform  \eqref{type_I_P_1_expanded} into  an equivalent form with only a finite number of  LMIs by applying the following lemma:

\textbf{\emph{Lemma 2:}} (\textit{General sign-definiteness} \cite{Gharavol2013sign}) Let ${\bf{Q}} \succeq \sum\limits_{i = 1}^{{\rm{I}}} {\left( {{\bf{A}}_i^H{{\bf{X}}_i}{{\bf{B}}_i} + {\bf{B}}_i^H{\bf{X}}_i^H{{\bf{A}}_i}} \right)} $ and ${\left\| {{{\bf{X}}_i}} \right\|_F} \le {\varepsilon _i}$, where ${\bf{Q}}={\bf{Q}}^H$. The condition  $\left\{ {{{\left\| {{{\bf{X}}_i}} \right\|}_F} \le {\varepsilon _i}} \right\}_{i = 1}^{{\rm{I}}} \Rightarrow {\bf{Q}} \succeq \sum\limits_{i = 1}^{{\rm{I}}} {\left( {{\bf{A}}_i^H{{\bf{X}}_i}{{\bf{B}}_i} + {\bf{B}}_i^H{\bf{X}}_i^H{{\bf{A}}_i}} \right)} $ holds if and only if there exist ${{\bar \lambda }_i} \ge 0,i \in \left\{ {1, \ldots ,{\rm{I}}} \right\}$ such that 
\begin{align}
\left[ {\begin{array}{*{20}{c}}
	{{\bf{Q}} - \sum\limits_{i = 1}^{{\rm{I}}} {{{\bar \lambda }_i}} {\bf{B}}_i^H{{\bf{B}}_i}}&{ - {\varepsilon _1}{\bf{A}}_1^H}& \cdots &{ - {\varepsilon _{{\rm{I}}}}{\bf{A}}_{{\rm{I}}}^H}\\
	{ - {\varepsilon _1}{{\bf{A}}_1}}&{{{\bar \lambda }_1}{\bf{I}}}& \cdots &{\bf{0}}\\
	\vdots & \vdots & \ddots & \vdots \\
	{ - {\varepsilon _{{\rm{I}}}}{{\bf{A}}_{{\rm{I}}}}}&{\bf{0}}& \cdots &{{{\bar \lambda }_{{\rm{I}}}}{\bf{I}}}
	\end{array}} \right] \succeq {\bf{0}}.
\end{align}
Based on Lemma 2, \eqref{type_I_P_1_expanded} can be  written as \eqref{const1_2_equi} (at the top of the next page), where ${{{\bar \lambda }_{1,k}}}\ge0$ and ${{{\bar \lambda }_{2,k}}}\ge0$  denote the corresponding  auxiliary variables.
\newcounter{mytempeqncnt0}
\begin{figure*}%¹«Ê½Î»ÖÃ°´Í¼·ÅÖÃµ÷Õû
	\normalsize
	\setcounter{mytempeqncnt0}{\value{equation}}
	\begin{align}
\left[ {\begin{array}{*{20}{c}}
	{{\beta _{c,k}} - \sigma _k^2 - {{\bar \lambda }_{1,k}} - {{\bar \lambda }_{2,k}}M}&{\left( {{{\bf{v}}^H}{{{\bf{\hat F}}}_k} + {\bf{\hat h}}_{d,k}^H} \right){{\bf{W}}_{ - k}}}&{{{\bf{0}}_{1 \times N}}}&{{{\bf{0}}_{1 \times N}}}\\
	{{\bf{W}}_{ - k}^H\left( {{\bf{\hat F}}_k^H{\bf{v}} + {{{\bf{\hat h}}}_{d,k}}} \right)}&{{{\bf{I}}_{K - 1 + N}}}&{{-\varepsilon _{d,k}}{\bf{W}}_{ - k}^H}&{{-\varepsilon _k}{\bf{W}}_{ - k}^H}\\
	{{{\bf{0}}_{N \times 1}}}&{{-\varepsilon _{d,k}}{{\bf{W}}_{ - k}}}&{{{\bar \lambda }_{1,k}}{{\bf{I}}_N}}&{{{\bf{0}}_N}}\\
	{{{\bf{0}}_{N \times 1}}}&{{-\varepsilon _k}{{\bf{W}}_{ - k}}}&{{{\bf{0}}_N}}&{{{\bar \lambda }_{2,k}}{\bf{I}}_N}
	\end{array}} \right] \succ {{\bf{0}}_{K + 3N}}, k \in {\cal K},\label{const1_2_equi}
	\end{align}
	\hrulefill % %ÕâÀïÓÐÒ»ÌõÏß£¬Èç¹ûÄãÏëÒª
	\vspace*{4pt} %Áô¿Õ°×£¬¿É×Ô¼ºµ÷Õû
\end{figure*}
%\begin{align}
%\left[ {\begin{array}{*{20}{c}}
%	{{\beta _{c,k}} - \sigma _k^2 - {{\bar \lambda }_{1,k}} - {{\bar \lambda }_{2,k}}M}&{\left( {{{\bf{v}}^H}{{{\bf{\hat F}}}_k} + {\bf{\hat h}}_{d,k}^H} \right){{\bf{W}}_{ - k}}}&{{{\bf{0}}_{1 \times N}}}&{{{\bf{0}}_{1 \times N}}}\\
%	{{\bf{W}}_{ - k}^H\left( {{\bf{\hat F}}_k^H{\bf{v}} + {{{\bf{\hat h}}}_{d,k}}} \right)}&{{{\bf{I}}_{K - 1 + N}}}&{{-\varepsilon _{d,k}}{\bf{W}}_{ - k}^H}&{{-\varepsilon _k}{\bf{W}}_{ - k}^H}\\
%	{{{\bf{0}}_{N \times 1}}}&{{-\varepsilon _{d,k}}{{\bf{W}}_{ - k}}}&{{{\bar \lambda }_{1,k}}{{\bf{I}}_N}}&{{{\bf{0}}_N}}\\
%	{{{\bf{0}}_{N \times 1}}}&{{-\varepsilon _k}{{\bf{W}}_{ - k}}}&{{{\bf{0}}_N}}&{{{\bar \lambda }_{2,k}}{\bf{I}}_N}
%	\end{array}} \right] \succ {{\bf{0}}_{K + 3N}}, k \in {\cal K},\label{const1_2_equi}
%\end{align}
To handle the  uncertainty $\Delta {{\bf{F}}_r} \in  {{\bf{\cal F}}_r}$ in constraint \eqref{Imperfect_const2}, we introduce  auxiliary variables $\left\{ {{\beta _{r,k}}}\ge 0 \right\}$ satisfying ${\beta _{r,k}} = \sum\limits_{i \ne k}^K {{{\left| {{{\bf{g}}^H}{{\bf{w}}_{c,i}}} \right|}^2}}  + \sum\limits_{n = 1}^N {{{\left| {{{\bf{g}}^H}{{\bf{w}}_{r,n}}} \right|}^2}}  + \sigma _t^2,k \in {\cal K}$, and  constraint \eqref{Imperfect_const2} can be then equivalently transformed as
\begin{align}
&{\left| {{{\bf{g}}^H}{{\bf{w}}_{c,k}}} \right|^2} \le {\beta _{r,k}}{r_{e,k,{\rm{th}}}}, {\theta _h} \in {\Phi _h},{\varphi _v} \in {\Phi _v},\notag\\
&\qquad\qquad\qquad\quad\qquad\qquad\qquad\qquad\Delta {{\bf{F}}_r} \in {{\bf{\cal F}}_r}, k\in{\cal K}, \label{everarop_1}\\
&\sum\limits_{i \ne k}^K {{{\left| {{{\bf{g}}^H}{{\bf{w}}_{c,i}}} \right|}^2}}  + \sum\limits_{n = 1}^N {{{\left| {{{\bf{g}}^H}{{\bf{w}}_{r,n}}} \right|}^2}}  + \sigma _t^2 \ge {\beta _{r,k}},\notag\\
&\qquad\qquad\qquad\quad {\theta _h} \in {\Phi _h},{\varphi _v} \in {\Phi _v}, \Delta {{\bf{F}}_r} \in {{\bf{\cal F}}_r}, k\in{\cal K}. \label{everarop_2}
\end{align}
Similar to the constraint \eqref{type_I_P_const1_new2}, we first   transform the inequalities  in \eqref{everarop_1} into   LMIs by applying   Schur’s complement, which yields  
\begin{align}
&\left[ {\begin{array}{*{20}{c}}
	{{\beta _{r,k}}{r_{e,k,{\rm{th}}}}}&{{{\bf{g}}^H}{{\bf{w}}_{c,k}}}\\
	{{\bf{w}}_{c,k}^H{\bf{g}}}&1
	\end{array}} \right] \succeq {{\bf{0}}_{2 }}, {\theta _h} \in {\Phi _h},{\varphi _v} \in {\Phi _v}, \notag\\
&\qquad\qquad\qquad\quad\qquad\qquad\qquad\qquad\Delta {{\bf{F}}_r} \in {{\bf{\cal F}}_r}, k \in {\cal K}. \label{everarop_1_S}
\end{align}
Recalling that ${{\bf{g}}^H} = {{\bf{v}}^H}{{\bf{F}}_r}$ and substituting ${{\bf{F}}_r}{\rm{ = }}{{{\bf{\hat F}}}_r} + \Delta {{\bf{F}}_r}$ into \eqref{everarop_1_S}, the following inequalities are obtained
\begin{align}
&\!\!\!\!\!\!\left[ {\begin{array}{*{20}{c}}
	{{\beta _{r,k}}{r_{e,k,{\rm{th}}}}}&{ {{\bf{v}}^H}{{{\bf{\hat F}}}_r}{{\bf{w}}_{c,k}}}\\
	{{\bf{w}}_{c,k}^H{\bf{\hat F}}_r^H{\bf{v}}}&1
	\end{array}} \right]\! +\! \left[ {\begin{array}{*{20}{c}}
	{{{\bf{v}}^H}}\\
	{{{\bf{0}}_{1 \times M}}}
	\end{array}} \right]\Delta {{\bf{F}}_r}\left[ {\begin{array}{*{20}{c}}
	{{{\bf{0}}_{N \times 1}}}&{{{\bf{w}}_{c,k}}}
	\end{array}} \right]\notag\\
& + \left[ {\begin{array}{*{20}{c}}
	{{{\bf{0}}_{1 \times N}}}\\
	{{\bf{w}}_{c,k}^H}
	\end{array}} \right]\Delta {\bf{F}}_r^H\left[ {\begin{array}{*{20}{c}}
	{{\bf{v}}}&{{{\bf{0}}_{M \times 1}}}
	\end{array}} \right] \succeq {{\bf{0}}_{2 }},{\theta _h} \in {\Phi _h},{\varphi _v} \in {\Phi _v},\notag\\
&\qquad\qquad\qquad\quad\qquad\qquad\qquad\qquad\quad\qquad\Delta {{\bf{F}}_r} \in {{\bf{\cal F}}_r}. \label{everarop_1_S_infity}
\end{align}
Based on Lemma 2,  constraint \eqref{everarop_1_S_infity}  involving an infinite number of inequalities can be recast as  a finite number of LMIs given by
\begin{align}
&\left[ {\begin{array}{*{20}{c}}
	{{\beta _{r,k}}{r_{e,k,{\rm{th}}}} - \bar \lambda_k M}&{ {{\bf{v}}^H}{{{\bf{\hat F}}}_r}{{\bf{w}}_{c,k}}}&{{{\bf{0}}_{1 \times N}}}\\
	{{\bf{w}}_{c,k}^H{\bf{\hat F}}_r^H{\bf{v}}}&1&{{-\varepsilon _r}{\bf{w}}_{c,k}^H}\\
	{{{\bf{0}}_{N \times 1}}}&{{-\varepsilon _r}{{\bf{w}}_{c,k}}}&{\bar \lambda_k {{\bf{I}}_N}}
	\end{array}} \right] \succeq {{\bf{0}}_{N + 2}},\notag\\
&\qquad \qquad\qquad\qquad\qquad\qquad{\theta _h} \in {\Phi _h},{\varphi _v} \in {\Phi _v},k \in {\cal K},\label{const2_2_equi}
\end{align}
where $\bar \lambda_k \ge 0 $ represents the corresponding auxiliary variables.

Although  constraint  \eqref{everarop_2} is not convex w.r.t $\bf v$, the left-hand side of  \eqref{everarop_2} is a quadratic function of $\bf v$. Thus,  we can obtain the following lower bound for ${\left| {{{\bf{g}}^H}{{\bf{w}}_{p,i}}} \right|^2},p \in \left\{ {c,r} \right\},i \in {\cal K}\cup {\cal N}$ at  any point ${\bf v}^r$  
\begin{align}
{\left| {{{\bf{g}}^H}{{\bf{w}}_{p,i}}} \right|^2} &\ge  - {\left| {{{\bf{v}}^{r,H}}{{\bf{F}}_r}{{\bf{w}}_{p,i}^r}} \right|^2} \notag\\
& +2{\mathop{\rm Re}\nolimits} \left\{ {{{\left( {{{\bf{v}}^H}{{\bf{F}}_r}{{\bf{w}}_{p,i}}} \right)}^H}\left( {{{\bf{v}}^{r,H}}{{\bf{F}}_r}{{\bf{w}}_{p,i}^r}} \right)} \right\}. \label{SINR_radar_1}
\end{align}
Substituting ${{\bf{F}}_r} = {{{\bf{\hat F}}}_r} + \Delta {{\bf{F}}_r}$ into ${\left| {{{\bf{v}}^{r,H}}{{\bf{F}}_r}{{\bf{w}}_{p,i}^r}} \right|^2}$, we have 
\begin{align}
&{\left| {{{\bf{v}}^{r,H}}{{{\bf{\hat F}}}_r}{{\bf{w}}_{p,i}^r} + {{\bf{v}}^{r,H}}\Delta {{\bf{F}}_r}{{\bf{w}}_{p,i}^r}} \right|^2} = {\left| {{{\bf{v}}^{r,H}}{{{\bf{\hat F}}}_r}{{\bf{w}}_{p,i}^r}} \right|^2} +\notag\\
&  {\rm{ve}}{{\rm{c}}^H}\left( {\Delta {\bf{F}}_r^*} \right){\left( {{{\left( {{{\bf{w}}_{p,i}^r}{\bf{w}}_{p,i}^{r,H}} \right)}^T} \otimes \left( {{{\bf{v}}^r}{{\bf{v}}^{r,H}}} \right)} \right)^T}{\rm{vec}}\left( {\Delta {\bf{F}}_r^*} \right)+\notag\\
& 2{\mathop{\rm Re}\nolimits} \left\{ {{{\bf{v}}^{r,H}}{{{\bf{\hat F}}}_r}{{\bf{w}}_{p,i}^r}\left( {{\bf{w}}_{p,i}^{r,H} \otimes {{\bf{v}}^{r,T}}} \right){\rm{vec}}\left( {\Delta {\bf{F}}_r^ * } \right)} \right\}.\label{SINR_radar_2}
\end{align}
In addition, substituting ${{\bf{F}}_r} = {{{\bf{\hat F}}}_r} + \Delta {{\bf{F}}_r}$ into ${{{\left( {{{\bf{v}}^H}{{\bf{F}}_r}{{\bf{w}}_{p,i}}} \right)}^H}\left( {{{\bf{v}}^{r,H}}{{\bf{F}}_r}{{\bf{w}}_{p,i}^r}} \right)}$, we have
\begin{align}
&\!\!{\left( {{{\bf{v}}^H}{{\bf{F}}_r}{{\bf{w}}_{p,i}}} \right)^H}\left( {{{\bf{v}}^{r,H}}{{\bf{F}}_r}{\bf{w}}_{p,i}^r} \right) \!= \!{\bf{w}}_{p,i}^H{\bf{\hat F}}_r^H{\bf{v}}{{\bf{v}}^{r,H}}{{{\bf{\hat F}}}_r}{\bf{w}}_{p,i}^r +{\bf{w}}_{p,i}^H\times \notag\\
&\!\!{\bf{\hat F}}_r^H{\bf{v}}{\rm{ve}}{{\rm{c}}^H}\left( {\Delta {\bf{F}}_r^*} \right)\left( {{\bf{w}}_{p,i}^r \otimes {{\bf{v}}^{r,*}}} \right) + {{\bf{v}}^{r,H}}{{{\bf{\hat F}}}_r}{\bf{w}}_{p,i}^r\left( {{\bf{w}}_{p,i}^H \otimes {{\bf{v}}^T}} \right)\times\notag\\
&\!\!{\rm{vec}}\left( {\Delta {\bf{F}}_r^*} \right)\! +\! {\rm{ve}}{{\rm{c}}^H}\left( {\Delta {\bf{F}}_r^*} \right)\left( {\left( {{\bf{w}}_{p,i}^r{\bf{w}}_{p,i}^H} \right) \otimes \left( {{{\bf{v}}^{r,*}}{{\bf{v}}^T}} \right)} \right){\rm{vec}}\left( {\Delta {\bf{F}}_r^*} \right).\label{SINR_radar_3}
\end{align}
Based on \eqref{SINR_radar_1}, \eqref{SINR_radar_2}, and \eqref{SINR_radar_3}, a lower bound for constraint \eqref{everarop_2} is given by
\begin{align}
&\!\!\!\!\!\!{\rm{ve}}{{\rm{c}}^H}\left( {\Delta {\bf{F}}_r^*} \right){{\bf{H}}_{{\rm{temp}}}}{\rm{vec}}\left( {\Delta {\bf{F}}_r^*} \right) +  \left( {\sum\limits_{i \ne k}^K {{c_{c,i}} + \sum\limits_{n = 1}^N {{c_{r,n}}} } } \right) + \sigma _t^2+\notag\\
& 2{\mathop{\rm Re}\nolimits} \left\{ {\left( {\sum\limits_{i \ne k}^K {{\bf{\hat g}}_{c,i}^H + \sum\limits_{n = 1}^N {{\bf{\hat g}}_{r,i}^H} } } \right){\rm{vec}}\left( {\Delta {\bf{F}}_r^*} \right)} \right\} \ge {\beta _{r,k}}, {\theta _h} \in {\Phi _h},\notag\\
&\qquad\qquad\qquad\quad\qquad\qquad {\varphi _v} \in {\Phi _v},\Delta {{\bf{F}}_r} \in {{\bf{\cal F}}_r}, k\in{\cal K},\label{SINR_radar_4}
\end{align}
where ${{\bf{H}}_{-k}} = \sum\nolimits_{i \ne k}^K {\left( {{{{\bf{\bar H}}}_{c,i}} + {\bf{\bar H}}_{c,i}^H - {\bf{\bar H}}_{c,i}^r} \right)}  + \sum\nolimits_{n = 1}^N {\left( {{{{\bf{\bar H}}}_{r,n}} + {\bf{\bar H}}_{r,n}^H - {\bf{\bar H}}_{r,n}^r} \right)} $, ${{{\bf{\bar H}}}_{p,i}} = \left( {{\bf{w}}_{p,i}^r{\bf{w}}_{p,i}^H} \right) \otimes \left( {{{\bf{v}}^{r,*}}{{\bf{v}}^T}} \right)$,
${\bf{\bar H}}_{p,i}^r = \left( {{\bf{w}}_{p,i}^r{\bf{w}}_{p,i}^{r,H}} \right) \otimes {\left( {{{\bf{v}}^r}{{\bf{v}}^{r,H}}} \right)^T}$, ${\bf{\hat g}}_{p,i}^H = {{\bf{v}}^H}{{{\bf{\hat F}}}_r}{{\bf{w}}_{p,i}}\left( {{\bf{w}}_{p,i}^{r,H} \otimes {{\bf{v}}^{r,T}}} \right) + {{\bf{v}}^{r,H}}{{{\bf{\hat F}}}_r}{\bf{w}}_{p,i}^r\left( {{\bf{w}}_{p,i}^H \otimes {{\bf{v}}^T}} \right) - {{\bf{v}}^{r,H}}{{{\bf{\hat F}}}_r}{\bf{w}}_{p,i}^r\left( {{\bf{w}}_{p,i}^{r,H} \otimes {{\bf{v}}^{r,T}}} \right)$, and ${c_{p,i}} = 2{\mathop{\rm Re}\nolimits} \left\{ {{\bf{w}}_{p,i}^H{\bf{\hat F}}_r^H{\bf{v}}{{\bf{v}}^{r,H}}{{{\bf{\hat F}}}_r}{\bf{w}}_{p,i}^r} \right\}{\rm{ - }}{\left| {{{\bf{v}}^{r,H}}{{{\bf{\hat F}}}_r}{\bf{w}}_{p,i}^r} \right|^2}$. Thus, based  on Lemma 1, constraint \eqref{SINR_radar_4} can be transformed  to a finite number of  LMIs given by 
\begin{align}
&\!\!\!\!\!\!\!\!\left[ {\begin{array}{*{20}{c}}
	{{{\bf{H}}_{-k}} + {\lambda _{r,k}}{{\bf{I}}_{MN}}}&{{{\left( {\sum\nolimits_{i \ne k}^K {{\bf{\hat g}}_{c,i}^H +   \sum\nolimits_{n = 1}^N {{\bf{\hat g}}_{r,i}^H} } } \right)}^H}}\\
	{\sum\nolimits_{i \ne k}^K {{\bf{\hat g}}_{c,i}^H + \sum\nolimits_{n = 1}^N {{\bf{\hat g}}_{r,i}^H} } }&{\left( {\sum\nolimits_{i \ne k}^K {{c_{c,i}} + \sum\nolimits_{n = 1}^N {{c_{r,n}}} } } \right) + c_{0,k}}
	\end{array}} \right] \notag\\
&\qquad\qquad\quad\qquad\succeq {{\bf{0}}_{MN + 1}}, {\theta _h} \in {\Phi _h},{\varphi _v} \in {\Phi _v},k \in {\cal K},\label{const2_2_eqv}
\end{align}
where $c_{0,k}=\sigma _t^2 - {\beta _{r,k}}-{\lambda _{r,k}}\varepsilon _r^2$ and ${\lambda _{r,k}} \ge 0$ denote  the corresponding auxiliary variables.

As a result,  problem \eqref{Imperfect_P} can be recast as 
  \begin{subequations} \label{Imperfect_P_equi}
	\begin{align}
	&\mathop {\max }\limits_{\left\{ {{{\bf{w}}_{c,k}}} \right\},\left\{ {{{\bf{w}}_{r,n}}} \right\},\left\{ {{v_m}} \right\},\left\{ {{\beta _{c,k}}} \right\},\bar \lambda ,\left\{ {{{\bar \lambda }_{1,k}},{{\bar \lambda }_{2,k}},{\lambda _{1,k}},{\lambda _{2,k}}},{\lambda _{r,k}} \right\},\chi } \chi  \label{Imperfect_P_equi_obj}\\
	& {\rm s.t.}~{{\bf{g}}^H}\left( {\sum\limits_{k = 1}^K {{{\bf{w}}_{c,k}}{\bf{w}}_{c,k}^H}  + \sum\limits_{n = 1}^N {{{\bf{w}}_{r,n}}{\bf{w}}_{r,n}^H} } \right){\bf{g}} \ge \chi ,{\theta _h} \in {\Phi _h},\notag\\
	&\qquad\qquad\qquad\qquad\qquad\quad\qquad{\varphi _v} \in {\Phi _v}, \Delta {{\bf{F}}_r} \in {{\bf{\cal F}}_r},\label{Imperfect_P_equi_const1}\\
	&\qquad \eqref{Imperfect_const3}, \eqref{Imperfect_const4}, \eqref{const1_1_eqv},\eqref{const1_2_equi}, \eqref{const2_2_equi}, \eqref{const2_2_eqv}.
	\end{align}
\end{subequations}
Similarly,  by applying Lemma 1, constraint \eqref{Imperfect_P_equi_const1} can be recast as 
\begin{align}
&\left[ {\begin{array}{*{20}{c}}
	{{{{\bf{\bar H}}}_{{\rm{temp}}}} + {{\tilde \lambda }_r}{{\bf{I}}_{MN}}}&{{{\left( {\sum\limits_{i{\rm{ = }}1}^K {{\bf{\hat g}}_{c,i}^H + \sum\limits_{n = 1}^N {{\bf{\hat g}}_{r,i}^H} } } \right)}^H}}\\
	{\sum\limits_{i{\rm{ = }}1}^K {{\bf{\hat g}}_{c,i}^H + \sum\limits_{n = 1}^N {{\bf{\hat g}}_{r,i}^H} } }&{\left( {\sum\limits_{i \ne k}^K {{c_{c,i}} + \sum\limits_{n = 1}^N {{c_{r,i}}} } } \right){\rm{ - }}\chi-{{\tilde \lambda} _r}\varepsilon _r^2}
	\end{array}} \right] \notag\\
&\qquad\qquad\qquad\qquad\qquad\succeq {{\bf{0}}_{MN + 1}},{\theta _h} \in {\Phi _h},{\varphi _v} \in {\Phi _v}, \label{Imperfect_P_equi_const1_equoi}
\end{align}
where ${{{\bf{\bar H}}}_{{\rm{temp}}}} = \sum\nolimits_{i = 1}^K {\left( {{{{\bf{\bar H}}}_{c,i}} + {\bf{\bar H}}_{c,i}^H - {\bf{\bar H}}_{c,i}^r} \right)}  + \sum\nolimits_{n = 1}^N {\left( {{{{\bf{\bar H}}}_{r,n}} + {\bf{\bar H}}_{r,n}^H - {\bf{\bar H}}_{r,n}^r} \right)} $ and  ${{\tilde \lambda} _r} $ is the auxiliary variable.
To solve problem \eqref{Imperfect_P_equi}, an AO algorithm is proposed to alternatively optimize transformers and IRS phase shifts until  convergence is reached. Below, we elaborate on how to solve these two subproblems.
\subsubsection{Optimizing BS beamformers with fixed IRS phase shifts}  This subproblem is given by 
  \begin{subequations} \label{Imperfect_P_equi_sub1}
	\begin{align}
	&\mathop {\max }\limits_{\left\{ {{{\bf{w}}_{c,k}},{{\bf{w}}_{r,n}},{\beta _{c,k}},{{\bar \lambda }_k},{{\bar \lambda }_{1,k}},{{\bar \lambda }_{2,k}},{\lambda _{1,k}},{\lambda _{2,k}},{\lambda _{r,k}}} \right\},\chi } \chi  \label{Imperfect_P_equi_sub1_obj}\\
	& {\rm s.t.}~\eqref{Imperfect_const3},  \eqref{const1_1_eqv},\eqref{const1_2_equi}, \eqref{const2_2_equi}, \eqref{const2_2_eqv},\eqref{Imperfect_P_equi_const1_equoi}.
	\end{align}
\end{subequations}
It can be  readily verified that problem  \eqref{Imperfect_P_equi_sub1} is a semi-definite program  (SDP), which   can be efficiently tackled by standard convex optimization  solvers.

\subsubsection{Optimizing IRS phase shifts with fixed BS beamformers} This subproblem is written as 
  \begin{subequations} \label{Imperfect_P_equi_sub2}
	\begin{align}
	&\mathop {\max }\limits_{\left\{ {{v_m},{\beta _{c,k}},{{\bar \lambda }_k},{{\bar \lambda }_{1,k}},{{\bar \lambda }_{2,k}},{\lambda _{1,k}},{\lambda _{2,k}},{\lambda _{r,k}}} \right\},\chi } \chi  \label{Imperfect_P_equi_sub2_obj}\\
	& {\rm s.t.}~\eqref{Imperfect_const4}, \eqref{const1_1_eqv},\eqref{const1_2_equi}, \eqref{const2_2_equi}, \eqref{const2_2_eqv},\eqref{Imperfect_P_equi_const1_equoi}.
	\end{align}
\end{subequations}
It can be  observed that all constraints are convex except \eqref{Imperfect_const4} due to the unit-modulus  constraint, which is in general difficult to  tackle. Fortunately, by applying the square penalty approach \cite{8811616},  problem  \eqref{Imperfect_P_equi_sub2} is equivalent to
  \begin{subequations} \label{Imperfect_P_equi_sub2_equi}
	\begin{align}
	&\mathop {\max }\limits_{\left\{ {{v_m},{\beta _{c,k}}{{\bar \lambda }_k},{{\bar \lambda }_{1,k}},{{\bar \lambda }_{2,k}},{\lambda _{1,k}},{\lambda _{2,k}},{\lambda _{r,k}}} \right\},\chi } \chi  + \bar \rho {\left\| {\bf{v}} \right\|^2} \label{Imperfect_P_equi_sub2_equi_obj}\\
	& {\rm s.t.}~\left| {{v_m}} \right| \le 1, m \in {\cal M}, \label{Imperfect_P_equi_sub2_equi_const1}\\
	&\qquad\eqref{const1_1_eqv},\eqref{const1_2_equi}, \eqref{const2_2_equi}, \eqref{const2_2_eqv},\eqref{Imperfect_P_equi_const1_equoi},
	\end{align}
\end{subequations}
where ${\bar \rho }$ represents a sufficiently large positive penalty parameter used to make  constraint \eqref{Imperfect_P_equi_sub2_equi_const1} met with equality at the optimal solution. Note that  this
equivalence does not require gradually adjusting ${\bar \rho }$ as a large ${\bar \rho }$ suffices.   The  rigorous proof  can be  found in  \cite[Theorem 1]{8811616} for details.
 To tackle the non-convex objective function in \eqref{Imperfect_P_equi_sub2_equi}, 
a lower bound for ${\left\| {\bf{v}} \right\|^2}$ is obtained by applying the SCA. Specifically, for any given  ${{\bf{v}}^r}$, we have 
\begin{align}
{\left\| {\bf{v}} \right\|^2} \ge  - {\left\| {{{\bf{v}}^r}} \right\|^2} + 2{\mathop{\rm Re}\nolimits} \left\{ {{{\bf{v}}^H}{{\bf{v}}^r}} \right\}, \label{Imperfect_P_equi_sub2_equi_const2}
\end{align}
which is linear w.r.t. $\bf v$. 
\begin{algorithm}[!t]
	%\algsetup{linenosize=\normalsize}
	%\normalsize
	\caption{The AO algorithm  for solving  problem   \eqref{Imperfect_P}.}
	\label{alg3}
	\begin{algorithmic}[1]
		\STATE  \textbf{Initialize} ${v_m}$  and ${\varepsilon}$.
		\STATE  \textbf{repeat}
		\STATE  \qquad Update BS beamformers by solving problem \eqref{Imperfect_P_equi_sub1}.
		\STATE  \qquad Update IRS phase shifts by solving problem \eqref{Imperfect_P_equi_sub2_equi_1}.
		\STATE \textbf{until} the fractional increase of the objective value is less than ${\varepsilon}$.
	\end{algorithmic}
\end{algorithm}

As a result,  based on \eqref{Imperfect_P_equi_sub2_equi_const2} and dropping irrelevant terms,  problem \eqref{Imperfect_P_equi_sub2_equi} can be approximated as 
  \begin{subequations} \label{Imperfect_P_equi_sub2_equi_1}
	\begin{align}
	&\mathop {\max }\limits_{\left\{ {{v_m},{\beta _{c,k}},{{\bar \lambda }_k},{{\bar \lambda }_{1,k}},{{\bar \lambda }_{2,k}},{\lambda _{1,k}},{\lambda _{2,k}},{\lambda _{r,k}}} \right\},\chi } \chi {\rm{ + }}2\bar \rho {\rm{Re}}\left\{ {{{\bf{v}}^H}{{\bf{v}}^r}} \right\} \label{Imperfect_P_equi_sub2_equi_obj_1}\\
	& {\rm s.t.}~\eqref{const1_1_eqv},\eqref{const1_2_equi}, \eqref{const2_2_equi}, \eqref{const2_2_eqv},\eqref{Imperfect_P_equi_const1_equoi},\eqref{Imperfect_P_equi_sub2_equi_const1},
	\end{align}
\end{subequations}
which is convex and can be solved by convex optimization  solvers.

Finally, we alternately optimize the above two subproblems, and the details are  summarized in Algorithm~\ref{alg3}.  Since problems \eqref{Imperfect_P_equi_sub1}   and  \eqref{Imperfect_P_equi_sub2_equi_1} are SDPs, the  
 complexity of Algorithm~\ref{alg3}  is  given by ${\cal O}\left( {L\left( {K\left( {{{\left( {N + MN + 1} \right)}^{6.5}} + {{\left( {K + 3N} \right)}^{6.5}}} \right)\! + \!\bar K{{\left( {N + 2} \right)}^{6.5}}} \right.} \right.$
 \noindent $\left. {\left. { + \left( {\bar K + 1} \right){{\left( {MN + 1} \right)}^{6.5}}} \right)} \right)$, where $L$ stands for the number of iterations required for reaching convergence and ${\bar K}$ denotes the number of LMIs in  \eqref{const2_2_equi} and  \eqref{const2_2_eqv}.
\section{Numerical Results}
In this section, we provide numerical results    to validate   the  secure transmission  performance in the IRS-aided ISAC system. A three dimensional   coordinate setup measured in meters (m) is considered, where the  BS is located at $( 0,0,2.5)~\rm m$  and the users are uniformly and randomly distributed in a circle of a radius $2~\rm m$  centered at $\left( {20,5,0 } \right)~{\rm m}$, while the IRS is deployed  at $(20 , 0, 2.5 )~{\rm m}$.
The distance-dependent path loss model is given by $L\left( {\hat d} \right) = {c_0}{\left( {{\hat d}/{d_0}} \right)^{ - \hat \alpha }}$,
where ${c_0} =-30~{\rm dB}$ is the path loss at the reference distance $d_0=1$ m,  ${\hat d}$ is the link distance, and $\hat \alpha$ is the path loss exponent.   The  target is  located at   azimuth direction  $\theta  =  - 30^ \circ  $ and  elevation direction  $ \varphi  = 40^ \circ $.
We assume that the distance between the IRS and the target is $10~\rm m$ with a  path loss exponent of $2$, and 
assume that  the  BS-IRS link and  the IRS-user link    follow Rician fading with a Rician factor of $3~{\rm dB}$ and a  path loss exponent of $2.2$, 
while the  BS-user link  follows Rayleigh  fading with a  path loss exponent of $3.6$. 
The minimum communication SINR and the  maximum tolerable intercepting SINR  are assumed to be the same for all users, i.e., $r_{c,\rm th}=r_{k,\rm th}, {r_{e,{\rm{th}}}}{\rm{ = }}{r_{e,k,{\rm{th}}}}, k\in {\cal K}$. Unless otherwise specified, we set   $N=4$, $K=3$, $\theta  =  - 30^ \circ , \varphi  = 40^ \circ $,  ${\sigma_t ^2}={\sigma_k ^2}{\rm{ =  - }}90~{\rm dBm},\forall k$,    $\rho=0.1$, $c=0.85$, $\varepsilon_{\rm in}=10^{-2}$, and ${\varepsilon}=\varepsilon_{\rm out}=10^{-4}$.
\subsection{Perfect CSI and Known  Target Location}
In this subsection, we consider the ideal case where the   CSI and the target  location are known at the BS, and the  penalty-based algorithm, i.e., Algorithm 1, is employed.
	\begin{figure}[!t]
	\centerline{\includegraphics[width=3.5in]{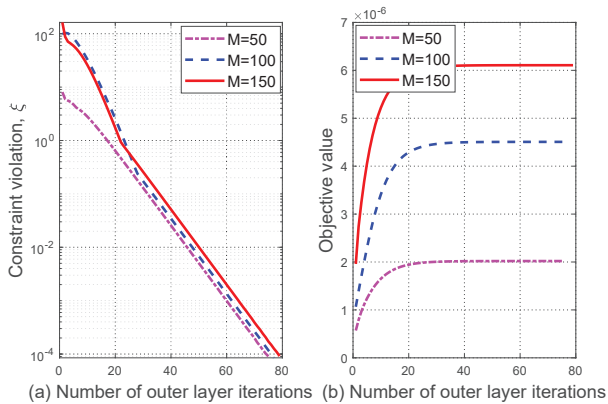}}
	\caption{Convergence behaviour of Algorithm~\ref{alg1} under $P_{\max}=40~{\rm dBm}$, $r_{c,\rm th}=10~{\rm dB}$, and  $r_{e,\rm th}=0~{\rm dB}$.} \label{figPenaltyconvergence}
%	\vspace{-0.5cm}
\end{figure}
\subsubsection{Convergence Behavior of Algorithm~\ref{alg1}}
We first study the convergence behavior of Algorithm~\ref{alg1} for different numbers of  IRS reflecting elements, namely $M=50$, $M=100$, and $M=150$,  as shown in Fig.~\ref{figPenaltyconvergence}. 
 It  is observed  from Fig.~\ref{figPenaltyconvergence}(a) that the constraint violation parameter $\xi$  converges   very rapidly to  a predefined accuracy $10^{-4}$ after about  $75$-$80$ iterations for all values of $M$.   Note that the predefined accuracy value of $10^{-4}$ is sufficiently small for ensuring that   constraint  
 \eqref{type_I_P_new_const3}  is essentially  met with equality at the optimal solution, since we normalize the channel coefficients by the noise power so that the auxiliary variables are inherently large to guarantee sufficient numerical accuracy. To see it more clearly, we can observe  from Fig.~\ref{figPenaltyconvergence}(b) that the objective value of problem \eqref{type_I_P_new_penbalty} converges quickly for different $M$,  which   demonstrates the  efficiency of Algorithm~\ref{alg1}.

To show the superiority of the proposed approach, we consider the  following  approaches for comparison. 
\begin{itemize}
	\item \textbf{Proposed approach:} This is our proposed approach  described in  Algorithm~\ref{alg1} in Section III. 
	\item \textbf{Communication  signal only:} Similar to the proposed approach, but without dedicated radar waveforms.
	\item \textbf{Separate beamforming:} This approach optimizes the transmit beamformers and IRS phase shifts separately. The algorithm first obtains  the IRS phase-shift matrix  by  maximizing the norm of the IRS's cascaded channel towards the desired sensing target, i.e, $\mathop {\max }\limits_{\bf{\Theta }} \left\| {{\bf{g}}_r^H{\bf{\Theta G}}} \right\|$. Then, with the obtained 
	${\bf{\Theta }}$, the transmit beamformers are obtained by solving problem \eqref{type_I_P}. 
	\item \textbf{Communication-based zero-forcing (ZF):} The IRS phase-shift matrix is obtained in the   same way as the  separate beamforming  approach, while the communication beamformers, ${{{\bf{w}}_{c,k}}}, k \in {\cal K}$, are forced to lie in the null space of the target's channel, i.e., ${{\bf{g}}^H}{{\bf{w}}_{c,k}}=0,k \in {\cal K}$. The communication covariance matrices are given by  ${{\bf{W}}_{c,k}} = \left( {{{\bf{I}}_N} - {\bf{g}}{{\bf{g}}^H}/{{\left\| {\bf{g}} \right\|}^2}} \right){{{\bf{\hat W}}}_{c,k}}{\left( {{{\bf{I}}_N} - {\bf{g}}{{\bf{g}}^H}/{{\left\| {\bf{g}} \right\|}^2}} \right)^H}$, where ${\rm{rank}}\left( {{{{\bf{\hat W}}}_{c,k}}} \right) = 1,{{{\bf{\hat W}}}_{c,k}} \succeq {{\bf{0}}_N}$. Then,  ${{{\bf{\hat W}}}_{c,k}}$ and the radar covariance matrices  are jointly optimized by using the  AO algorithm. 
	\item \textbf{Sensing-based ZF:} Similar to the communication-based ZF approach,  the radar beamformers, i.e., ${{{\bf{w}}_{r,n}}},n\in {\cal N}$, are   forced to lie in the null space of the users' channels i.e., ${{\bf{h}}_k^H{{\bf{w}}_{r,n}}}=0, k\in {\cal K}, n\in {\cal N}$. The radar beamformers are designed as  ${{\bf{W}}_r} = {\bf{V}}{{{\bf{\hat W}}}_r}$, where ${\bf{V}}$ represents the  last $N-K$ right singular vectors of ${\bf{H}} = {\left[ {{{\bf{h}}_1}, \ldots ,{{\bf{h}}_K}} \right]^H}$. Then, the communication beamformers and ${{{\bf{\hat W}}}_r}$ are jointly optimized by using the  penalty-based algorithm.
	\item \textbf{Random phase:} The IRS phase shifts are generated randomly following a  uniform distribution over $\left[ {0,2\pi } \right)$.
\end{itemize}	
	\begin{figure}[!t]
	\centerline{\includegraphics[width=3.2in]{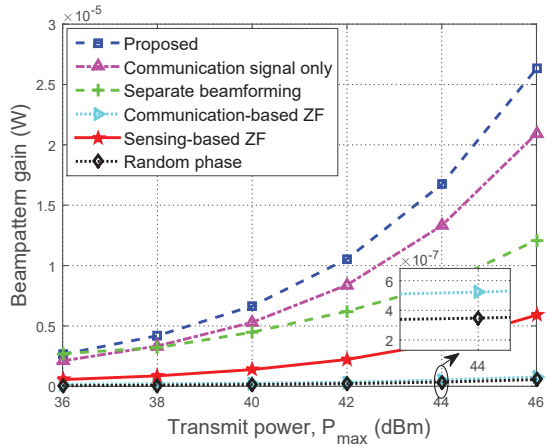}}
	\caption{Beampattern gain versus $P_{\max}$ under $M=100$, $r_{c,\rm th}=10~{\rm dB}$, and  $r_{e,\rm th}=0~{\rm dB}$.}\label{figVSPower}
%		\vspace{-0.5cm}
\end{figure}
%\begin{figure}[!t]
%	\centering
%	\begin{minipage}[t]{0.45\textwidth}
%		\centering
%		\includegraphics[width=3.5in]{Penalty_convergence.eps}
%		\caption{Convergence behaviour of Algorithm~\ref{alg1} under $P_{\max}=40~{\rm dBm}$, $r_{c,\rm th}=10~{\rm dB}$, and  $r_{e,\rm th}=0~{\rm dB}$.} \label{figPenaltyconvergence}
%	\end{minipage}
%	\hspace{10pt}
%	\begin{minipage}[t]{0.45\textwidth}
%		\centering
%		\includegraphics[width=2.9in]{vspower.eps}
%		\caption{Beampattern gain versus $P_{\max}$ under $M=100$, $r_{c,\rm th}=10~{\rm dB}$, and  $r_{e,\rm th}=0~{\rm dB}$.}\label{figVSPower}
%	\end{minipage}
%	\vspace{-10pt} 
%\end{figure}
\subsubsection{Beampattern Gain Versus Transmit Power} 
In Fig.~\ref{figVSPower}, we compare the beampattern gain of the above approaches  versus $P_{\max}$. We see that the beampattern gain  for all methods increases monotonically with  $P_{\max}$ since   the  co-channel  interference   is    suppressed and increasing the available power improves   the  beampattern gain. In addition, we observe that the proposed approach outperforms the ``Communication signal only'' case, which indicates  the benefit of dedicated radar signals. This  
can be explained as follows. The additional radar signals provide more DoFs for algorithm optimization, which  improves the system performance, and  to prevent the eavesdropping by the target,  more power must be allocated to the radar signals  and the beampattern gain is thus increased. 
Moreover, we observe that the beampattern gain obtained by  the approaches without IRS  phase shift optimization increases marginally as $P_{\max}$ increases since the signals reflected  by the IRS in this case are propagated in many random directions, thus results in a low received power level. Furthermore, compared to the ``Separate beamforming'', ``Communication-based ZF'', and ``Sensing-based ZF'' approaches, our proposed approach achieves significant beampattern gains, which illustrates the benefit of joint design of the  transmit beamformers and IRS phase shifts.

\subsubsection{Beampattern Gain Versus Number of IRS Reflecting Elements} 
\begin{figure}[!t]
 		\centerline{\includegraphics[width=3.2in]{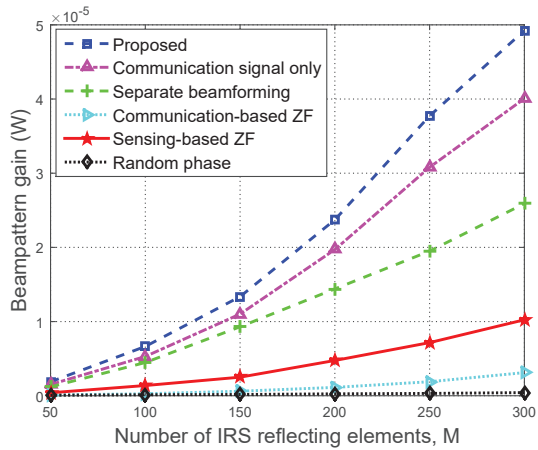}}
 		\caption{Beampattern gain versus $M$ under  $P_{\max}=40~{\rm dBm}$, $r_{c,\rm th}=10~{\rm dB}$, and  $r_{e,\rm th}=0~{\rm dB}$.} \label{figvsM}
% 		\vspace{-0.5cm}
 	\end{figure}

In Fig.~\ref{figvsM}, we compare the beampattern gain for all approaches versus $M$. 
It is observed that the proposed approach outperforms   the ``Random phase'' approach, and  the system performance gap is more pronounced  for a larger $M$.
 This is because installing more passive reflecting elements provides more DoFs for  resource allocation, which is beneficial for achieving higher
 beamforming gain, thereby improving  the beampattern gain when the IRS phase shifts are well adjusted.  In addition, we again observe that our proposed approach outperforms the use of only communication signals, further amplifying the benefit of using dedicated radar signals. Moreover, the performance gap between our proposed approach and  the ``Separate beamforming'', ``Communication-based ZF'', and ``Sensing-based ZF'' approaches is magnified as $M$ increases, which again demonstrates the benefit of joint design of the transmit beamformers and IRS phase shifts.  
% Furthermore, it is  clearly evident that our proposed algorithm is appealing for large-scale IRS with hundreds or even thousands of reflecting elements since each subproblem in Algorithm~\ref{alg1} is solved with low complexity, which demonstrates the superiority of Algorithm~\ref{alg1}.
 	\begin{figure}[!t]
 		\centerline{\includegraphics[width=3.2in]{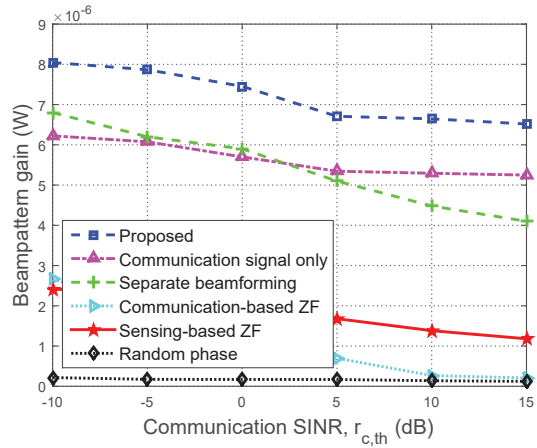}}
 		\caption{Beampattern gain versus $r_{c,\rm th}$ under  $M=100$,  $P_{\max}=40~{\rm dBm}$,  and  $r_{e,\rm th}=0~{\rm dB}$.}\label{figVSCommunicationSINR}
% 		\vspace{-0.5cm}
 	\end{figure}
% \begin{figure}[!t]
%	\centering
%	\begin{minipage}[t]{0.45\textwidth}
%		\centering
%		\includegraphics[width=3in]{vsM.eps}
%		\caption{Beampattern gain versus $M$ under  $P_{\max}=40~{\rm dBm}$, $r_{c,\rm th}=10~{\rm dB}$, and  $r_{e,\rm th}=0~{\rm dB}$.} \label{figvsM}
%	\end{minipage}
%	\hspace{10pt}
%	\begin{minipage}[t]{0.45\textwidth}
%		\centering
%		\includegraphics[width=3in]{vssuerSINR.eps}
%		\caption{Beampattern gain versus $r_{c,\rm th}$ under  $M=100$,  $P_{\max}=40~{\rm dBm}$,  and  $r_{e,\rm th}=0~{\rm dB}$.}\label{figVSCommunicationSINR}
%	\end{minipage}
%	\vspace{-10pt} 
%\end{figure}
\subsubsection{Beampattern Gain Versus Minimum SINR Required by Communication Users} 
In Fig.~\ref{figVSCommunicationSINR}, the achieved beampattern gain  is plotted versus the communication users' SINR requirement  $r_{c,{\rm th}}$. 
As expected, a more stringent QoS requirement for the users results in a lower beamforming gain to the target, since the BS and IRS must focus more energy towards the communication users. In addition, we observe that  the performance gap between our proposed approach and the  ``Separate beamforming''  approach becomes  smaller as   $r_{c,\rm th}$  decreases. This
is because    in this case the SINR at the users can
be easily satisfied, and thus extra   radar  and  communication power can be used to improve the beampatter gain.  Moreover, we observe that the performance of the  ``Communication-based ZF''  approach degrades quickly as  $r_{c,{\rm th}}$ increases. This is because    communication signals are forced to lie in the null space of target's channel, which indicates that no  user  information is leaked to the target and only the radar signals can be used to increase the beampattern gain. On the other hand, increasing the radar power potentially degrades the user SINR, which limits the improvement of beampattern gain. 
 	\begin{figure}[!t]
	\centerline{\includegraphics[width=3.2in]{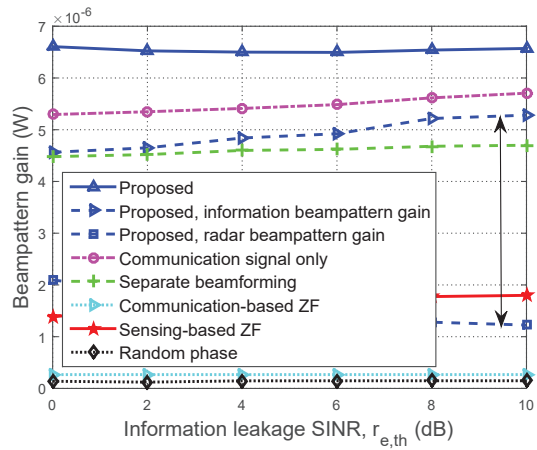}}
	\caption{Beampattern gain versus $r_{e,\rm th} $ under $M=100$,  $P_{\max}=40~{\rm dBm}$, and $r_{c,\rm th}=10~{\rm dB}$.}\label{figVSLeakageSINR}
%	\vspace{-0.5cm}
\end{figure} 
\subsubsection{Beampattern Gain Versus   Maximum    Information Leakage SINR to Target} 
 We further study the beampattern gain  versus the leakage constraint $r_{e,{\rm th}}$   in Fig.~\ref{figVSLeakageSINR}. Interestingly, we observe that the beampattern gain obtained by the proposed approach   remains nearly unchanged with $r_{e,{\rm th}}$.  To unveil the reason behind this, the separate radar  and communication power contributions to  the beampattern gain versus $r_{e,{\rm th}}$ are studied, i.e., $\sum\nolimits_{n = 1}^N {{{\left| {{\bf{w}}_{r,n}^H{\bf{g}}} \right|}^2}} $ and $\sum\nolimits_{k = 1}^K {{{\left| {{\bf{w}}_{c,k}^H{\bf{g}}} \right|}^2}} $, which correspond to the ``Proposed, radar beampattern gain''  and the  ``Proposed, information beampattern gain''  approaches, respectively.  We see that as the requirement on signal leakage to the target is made less stringent (i.e., $r_{e,{\rm th}}$ increases), less transmit power is allocated to radar signals to deteriorate the reception by the eavesdropping target, while more power is allocated to the information signals to improve the communication QoS. In the end, these two trends offset each other, and the sum of the two components results in a nearly constant beampattern gain. In addition, we observe that the performance gain obtained by the ``Sensing-based ZF''  approach increases marginally as $r_{e,\rm th} $ increases due to the limited DoFs available for design of the  radar beamformers.
 	\begin{figure}[!t]
	\centerline{\includegraphics[width=3.2in]{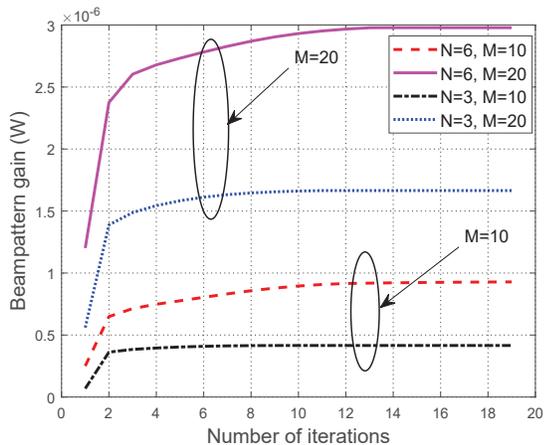}}
	\caption{Convergence behaviour of Algorithm~\ref{alg3} for different $M$ and $N$.}\label{Impefect_convergence}
%	\vspace{-0.5cm}
\end{figure}
%\begin{figure}[!t]
%	\centering
%	\begin{minipage}[t]{0.45\textwidth}
%		\centering
%		\includegraphics[width=3in]{vstargetSINR.eps}
%		\caption{Beampattern gain versus $r_{e,\rm th} $ under $M=100$,  $P_{\max}=40~{\rm dBm}$, and $r_{c,\rm th}=10~{\rm dB}$.}\label{figVSLeakageSINR}
%	\end{minipage}
%	\hspace{10pt}
%	\begin{minipage}[t]{0.45\textwidth}
%		\centering
%		\includegraphics[width=3in]{Impefect_convergence.eps}
%		\caption{Convergence behaviour of Algorithm~\ref{alg3} for different $M$ and $N$.}\label{Impefect_convergence}
%	\end{minipage}
%	\vspace{-10pt} 
%\end{figure}
\subsection{Imperfect CSI and Uncertain Target Location}
In this subsection, we consider the case with imperfect CSI  and   an  unknown target location, and we   propose  Algorithm~\ref{alg3}  to address the resulting problem. The azimuth and elevation target location ranges are set to
${\Phi _h} = \left[ { - {{35}^\circ }, - {{25}^\circ }} \right]$ and   ${\Phi _v} = \left[ {{{35}^\circ },{{45}^\circ }} \right]$, respectively.
We  define the relative amount of CSI errors    as ${{\hat \varepsilon }_r} = {\varepsilon _r}/{\left\| {\Delta {{\bf{F}}_r}} \right\|_F}$, ${{\hat \varepsilon }_k} = {\varepsilon _k}/{\left\| {\Delta {{\bf{F}}_k}} \right\|_F}$, and ${{\hat \varepsilon }_{d,k}} = {\varepsilon _{d,k}}/\left\| {\Delta {{\bf{h}}_{d,k}}} \right\|,\forall k$, respectively. For ease of exposition, we assume that all channels have the same level of CSI errors and define ${\varepsilon _{{\rm{error}}}} = {{\hat \varepsilon }_r} = {{\hat \varepsilon }_k} = {{\hat \varepsilon }_{d,k}}, \forall k$.

\subsubsection{Convergence Behavior of Algorithm~\ref{alg3}}In Fig.~\ref{Impefect_convergence}, the convergence behaviour of Algorithm~\ref{alg3} for different $M$ and $N$ under ${\varepsilon _{{\rm{error}}}}=0.01$, $P_{\max}=46~{\rm dBm}$, $K=2$,  $r_{c,\rm th}=10~{\rm dB}$, and  $r_{e,\rm th}=5~{\rm dB}$  is studied. It is observed that the obtained beampattern gain is monotonically increasing with the number of iterations and ultimately converges.  Even for  $M=20$ and $N=6$, the proposed algorithm converges in about 20 iterations, which demonstrates the effectiveness of Algorithm~\ref{alg3}. 
  %  \begin{figure}[!t]
%	\centerline{\includegraphics[width=3in]{convergence.eps}}
%	\caption{ Convergence behaviors of  the proposed MMSE-based algorithm and EW-based algorithm.} \label{convergence} 
%	\vspace{-0.5cm}
%\end{figure}
	\begin{figure}[!t]
		\centerline{\includegraphics[width=3.2in]{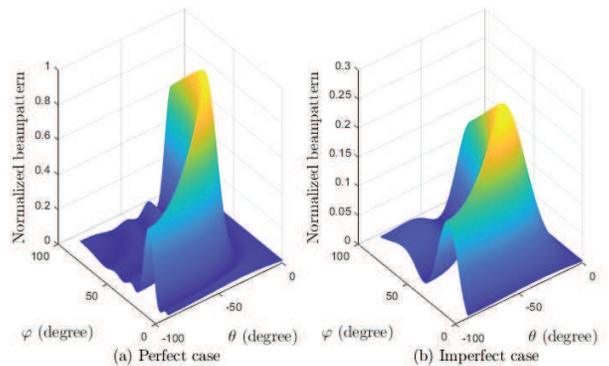}}
		\caption{Beampattern design for different system setups.}\label{BeampatternVSangle}
%		\vspace{-0.5cm}
	\end{figure}
%  \begin{figure}[!t]
%	\centering
%		\begin{minipage}[t]{0.45\textwidth}
%		\centering
%		\includegraphics[width=3.5in]{BeampatternVSangleV2.eps}
%		\caption{Beampattern design for different system setups.}\label{BeampatternVSangle}
%	\end{minipage}
%	\hspace{30pt}
%	\begin{minipage}[t]{0.45\textwidth}
%		\centering
%		\includegraphics[width=2.8in]{Impefect_VSM.eps}
%		\caption{Beampattern gain versus $M$ for different $N$ and  ${\varepsilon _{{\rm{error}}}}$.}\label{Impefect_VSM}
%	\end{minipage}
%	\vspace{-10pt} 
%\end{figure}
\subsubsection{Beampattern Design}
In Fig.~\ref{BeampatternVSangle}, we study the normalized beampattern obtained in the case with perfect CSI and the known target location and with the case of imperfect CSI and uncertain target location when
 $M=20$, $N=3$, ${\varepsilon _{{\rm{error}}}}=0.01$, $P_{\max}=46~{\rm dBm}$, $K=2$,  $r_{c,\rm th}=10~{\rm dB}$, and  $r_{e,\rm th}=5~{\rm dB}$. Both beampatterns are  normalized by the maximum value of these  two beampatterns.
It is observed that both of the beampatterns obtained by our proposed algorithms correctly focus their mainlobe towards the directions $\theta  =  - 30^ \circ$ and $ \varphi  = 40^ \circ $. In addition, we observe that both beampatterns  have    sidelobe regions due to the imposed SINR constraints for the users  as well as the information leakage to the eavesdropping target. Furthermore, we observe that the mainlobe in the imperfect CSI case is more flat and wide than that with  perfect CSI case. This is expected since although the exact target location is unknown, its range of possible locations is known, so that the probing power should uniformly  cover this  area rather than focusing on  a point in  one direction. Moreover, we  observe that the peak beampattern gain of the imperfect CSI case is lower than with in the perfect CSI case due to the reduced available information.
\subsubsection{Beampattern Gain Versus $M$}
	\begin{figure}[!t]
	\centerline{\includegraphics[width=3.2in]{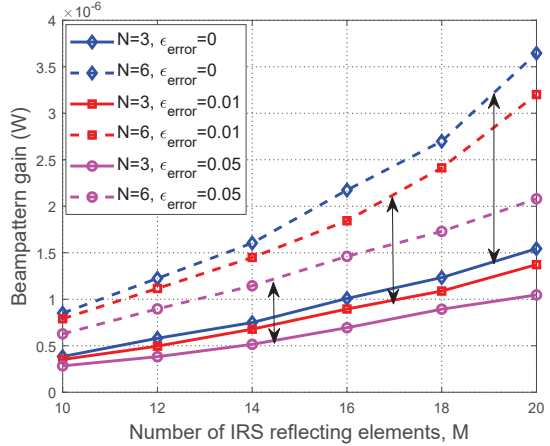}}
	\caption{Beampattern gain versus $M$ for different $N$ and  ${\varepsilon _{{\rm{error}}}}$.}\label{Impefect_VSM}
%	\vspace{-0.5cm}
\end{figure}
In Fig.~\ref{Impefect_VSM}, we study the beampattern gain versus $M$ for different $N$ and ${\varepsilon _{{\rm{error}}}}$  under   $P_{\max}=46~{\rm dBm}$, $K=2$,  $r_{e,\rm th}=5~{\rm dB}$, and $r_{c,\rm th}=10~{\rm dB}$.
A large ${\varepsilon _{{\rm{error}}}}$ indicates that the channel estimation error is magnified and ${\varepsilon _{{\rm{error}}}}=0$ corresponds to the  perfect CSI case. It is observed that the beampattern gain obtained by different  $N$ and ${\varepsilon _{{\rm{error}}}}$ monotonically  increases with $M$. This observation shows that by carefully designing the BS beamformers and the IRS phase shifts, the system performance can still be improved  with imperfect CSI even with  large channel estimation errors, e.g., ${\varepsilon _{{\rm{error}}}} = 0.05$.  
%In addition, we observe that the beampattern gain gap between the perfect CSI case and the imperfect CSI case becomes magnified as $M$ increases. This is because increasing $M$ exacerbates the resource allocation mismatch caused by the channel uncertainty.
Furthermore, we observe that for a fixed $M$,  the beampattern gain  increases with   $N$. This is due to the fact that more DoFs can be exploited for resource allocation to achieve higher array gain.

\section{Conclusion}
In this paper, we proposed the use of IRS  to achieve simultaneous secure communication and sensing in the presence of an eavesdropping target and multiple communication users. The communication beamformers, the radar beamformers, and the IRS phase shifts were jointly optimized to maximize the sensing  beampattern gain while satisfying the minimum SINR required by the users and secrecy constraint for the eavesdropping target.   For the first scenario where the  CSI of the user links and the   target location are known, a  penalty-based algorithm was proposed to solve the formulated non-convex optimization problem. In particular, the beamformers were obtained  via a semi-closed-form solution   using the  Lagrange duality method and the IRS phase shifts were obtained in   closed-form  by applying the MM method. For the second scenario where 
 the  CSI  and the   target location are imprecisely unknown, an efficient  AO algorithm   based on  the $\cal S$-procedure and sign-definiteness approaches
was proposed.  Simulation results  verified  the effectiveness of the proposed scheme in achieving a flexible trade-off between the communication quality  and the target  sensing quality and showed the capability   of the IRS for use in sensing and   improving the physical layer security  of ISAC systems.  In addition,   simulation results  also   illustrated   the  benefits of using  dedicated sensing signals   to improve  the sensing quality.

\bibliographystyle{IEEEtran}
\bibliography{SECIRS_TSP.bib}
\end{document}